\documentclass[graybox]{svmult}

\usepackage{aas_macros}
\usepackage{type1cm}        
\usepackage{makeidx}         
\usepackage{graphicx}        
\usepackage{multicol}        
\usepackage[bottom]{footmisc}

\usepackage{xspace}
\usepackage{newtxtext}       %
\usepackage{newtxmath}       
\usepackage{comment}
\usepackage{subcaption}
\usepackage[normalem]{ulem}
\def\msun{\rm M_{\odot}}

\def\xmm{{\em XMM-Newton}}

\def \xmm {{\em XMM-Newton}}

\def \ergsec{\hbox{erg s$^{-1}$}}

\def \hcm {\hbox {\ifmmode $ atom cm$^{-2}\else atom cm$^{-2}$\fi}}

\def\apjl{ApJ, }
\def\apjs{ApJS, }
\def \aap {A\&A, }

\def\ltsima{$\; \buildrel < \over \sim \;$}
\def\simlt{\lower.5ex\hbox{\ltsima}}
\def\gtsima{$\; \buildrel > \over \sim \;$}
\def\simgt{\lower.5ex\hbox{\gtsima}}

\def\psr1023{PSR J1023+0038}
\def\xss{XSS J12270-4859}
\def\3fgl{3FGL J1544.6-1125}
\def\j0838{3FGL J0838.8-2829}

\newcommand{\igr}{IGR~J18245--2452}

\makeindex             

\begin{document}

\title*{Transitional millisecond pulsars}
\author{Alessandro Papitto \& Domitilla de Martino}
\institute{Alessandro Papitto \at INAF -- Osservatorio Astronomico di Roma, via di Frascati 33, I-00078 Monte Porzio Catone (Roma), Italy, \email{alessandro.papitto@inaf.it} \and Domitilla de Martino \at INAF -- Osservatorio Astronomico di Capodimonte, Salita Moiariello 16, I-80131 Napoli, Italy, \email{domitilla.demartino@inaf.it}}
%
%
\maketitle

\abstract
{Millisecond pulsars in tight binaries have recently opened new challenges in our understanding of physical processes governing the evolution of binaries  and the interaction between astrophysical plasma and electromagnetic fields. Transitional systems that showed changes from rotation-powered to accretion powered states and vice versa have bridged the populations of radio and accreting millisecond pulsars, eventually demonstrating the tight evolutionary link envisaged by the recycling scenario.  A decade of discoveries and theoretical efforts have just grasped the complex phenomenology of transitional millisecond pulsars from the radio to the gamma-ray band. This review summarises the main properties of the three transitional millisecond pulsars discovered so far, as well as of candidates and related systems, discussing the various models proposed to cope with the multifaceted behaviour.}

\section{Introduction} 
\label{sec:1}

The observation of dramatic changes of state over a few weeks is what makes {\it transitional} a millisecond pulsar (MSP) in a binary system. These sources  experience  transitions from a rotation-powered regime,
in which they behave like radio pulsars whose wind prevents the in-fall of the matter lost by the low-mass  companion, to a regime in which they accrete and emit intense  high-energy radiation like X-ray binary systems, and vice versa. During the transitions the luminosity changes by at least an order of magnitude, likely due to variations in the mass inflow rate.   The discovery of transitional millisecond pulsars (tMSPs in the following) has been a key achievement in the investigation of the
evolution of MSPs. The recycling scenario
developed in the early 1980s had postulated that radio MSPs were spun
up in a previous Gyr-long phase of mass accretion
\cite{alp82,rad82,fab83,bha91}. It implied that these fast-spinning
neutron stars (NSs) were descendants of low-mass (donor mass
$<1\,M_{\odot}$) X-ray binaries (LMXBs). The discovery of radio MSPs in globular clusters \cite{lyn87}
and of accreting MSPs (AMXPs) in a handful of X-ray
transients \cite{wij98}  underpinned
this theory. Eventually, in 2009 a radio MSP binary was recognised to
have been previously surrounded by an accretion disc \cite{arc09}. A
few years later, the same binary and two more systems were surprisingly found to switch
back and forth accretion and rotation-powered regimes over much
shorter timescales than the secular recycling binary evolution
\cite{pap13,sta14,bas14}. 
Whether these MSPs represent an intermediate evolutionary stage before they
end as radio pulsars that completely devour their companions, or rather    experience  a distinct evolutionary path, has still to be
assessed. Certainly, they have provided us with a unique occasion to observe
the different possible outcomes of the interaction between a quickly
spinning magnetised NS and plasma lost by a companion star as they
unfold over timescales accessible to the human life.

TMSPs bridge a few classes of MSPs (see Sec.~\ref{sec:pop_msps}).  The interplay between the gravitational pull
exerted by the NS on the mass lost by the companion and the outward
pressure of the pulsar wind determines whether an MSP behaves either
as a rotation or as an accretion-powered source (see
Sec.~\ref{sec:changes}). As of June 2020, we currently know three transitional MSPs (see Sec.~\ref{sec:sources}). They have shown radio pulsar states (Sec.~\ref{sec:radiopsr}), a bright
accretion outburst in one of them (Sec.~\ref{sec:outburst}), but also  an enigmatic
X-ray {\it sub-luminous} disc state, which gave us a brand new view
of how LMXBs may behave at low mass accretion rates
(Sec.~\ref{sec:sublum}). The properties of tMSPs are so peculiar that
they are key signatures to identify  candidates that will likely
perform a transition in the future (Sec.~\ref{sec:cand}).  Like in
many cases, the discovery of tMSPs raised far more questions than it
answered. What makes a system transitional?  Are all MSPs in binaries
with an orbital period shorter than a day (both accreting and
eclipsing ones) transitional? How does the pulsar magnetic field
interact with the in-flowing gas? Are the rotation and accretion
powered regimes mutually exclusive, or do we rather see them mixed in
the sub-luminous state?  In Sec.~\ref{sec:models} we discuss current
models attempting to explain the enigmatic behaviour of these MSPs.

\section{The population of millisecond pulsar binaries}
\label{sec:pop_msps}

\begin{figure}[t!]
\includegraphics[scale=0.45]{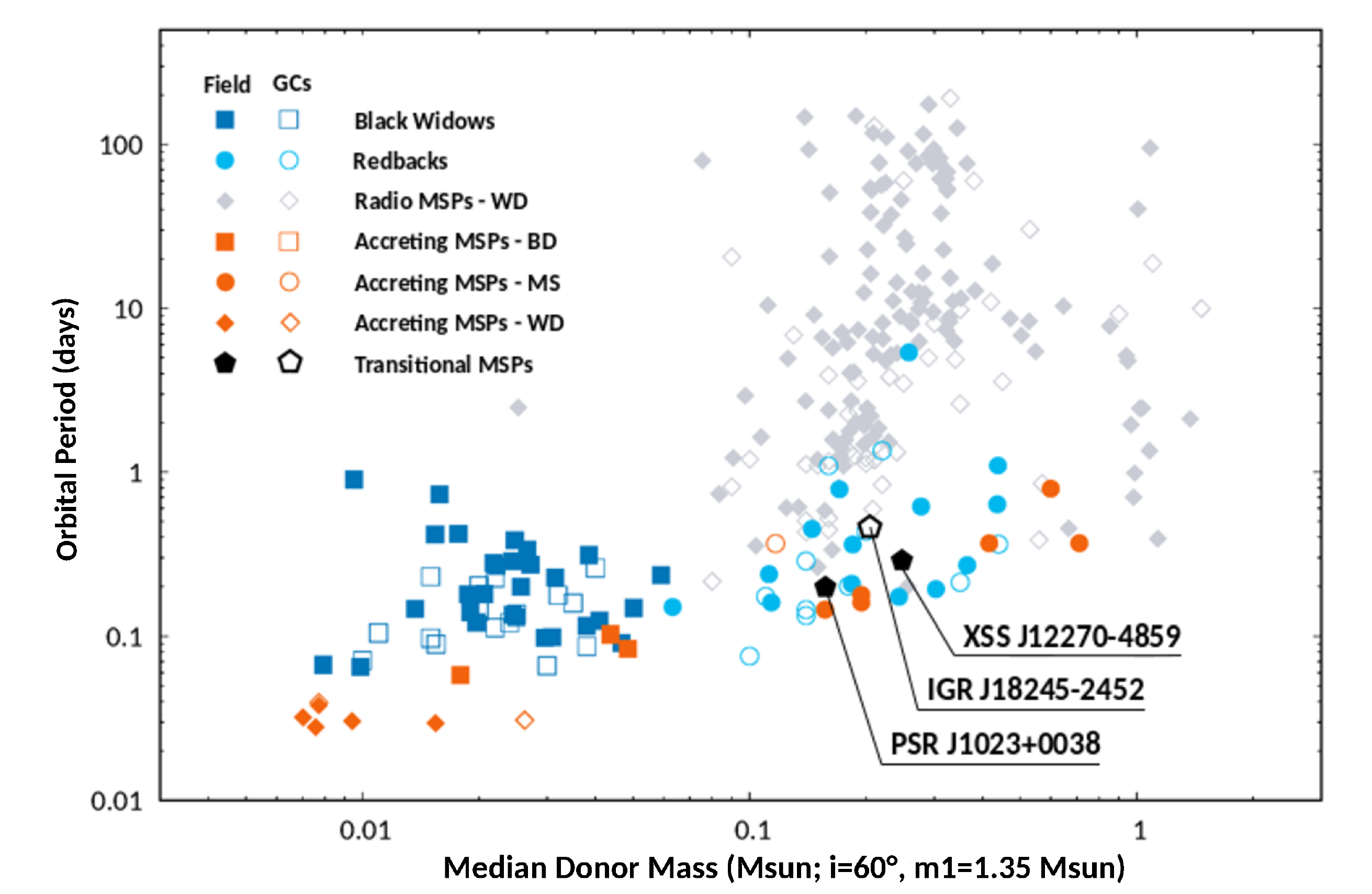}
\caption{Orbital period and median mass (evaluated from the pulsar
  timing parameters assuming an inclination of $60^{\circ}$ and an NS
  mass of 1.35~$\msun$) of black widows (blue squares), redbacks (cyan
  circles), non-interacting radio MSPs with a white dwarf companion
  (grey diamonds), AMXPs (orange symbols with shape depending on the
  donor type, squares for brown dwarfs, circles for main-sequence
  stars and diamonds for white dwarfs) and tMSPs (black
  pentagons). Filled and hollow symbols mark sources found in the
  Galactic field and globular clusters, respectively. }
\label{fig:psr}      
\end{figure}

As of June 2020, we know 20 accreting and 471 rotation-powered MSPs,
here defined by a spin period $< 30$~ms. Fig.~\ref{fig:psr} shows
the observed binary characteristics.

AMXPs  are all found in
X-ray transients, which undergo occasional outbursts reaching an X-ray
luminosity up to $\rm \sim 10^{36}-10^{38}\,erg\,s^{-1}$, interleaved
by long periods of quiescence ($\rm L_X \sim
10^{31}-10^{32}\,erg\,s^{-1}$; see \cite{pat12rev, cam18rev} for reviews). They are harboured in tight
($P_{orb}<1$~day) binary systems (see orange symbols in
Fig.~\ref{fig:psr}), and half of them showed thermonuclear type-I
bursts.  Only IGR\,J18245-2452 has been also
detected as a radio pulsar in quiescence, so far.

Rotation-powered radio MSPs  are much more
numerous. They amount to about 314 in the Galactic field\footnote{See
  the list maintained by D.~Lorimer \& E. Ferrara, available at
  \url{http://astro.phys.wvu.edu/GalacticMSPs/GalacticMSPs.txt}} and
157 in 30 globular clusters\footnote{See the list maintained
  P.~Freire, available at
  \url{http://www.naic.edu/~pfreire/GCpsr.html}}.  Here, we mainly
address the $\sim$100 MSPs in compact binaries ($P_{orb}<1$~day),
most of which show irregular radio eclipses due to the presence of
intrabinary material. Dubbed {\it "spiders"} \cite{rob13,str19}, these
binaries include {\it "black widows"} with very low-mass companions
(defined by $\rm < 0.06\,M_{\odot}$) and {\it "redbacks"} with a
hydrogen-rich secondary with a minimum mass of at least
$\approx$0.1$\msun$ (see Fig.~\ref{fig:psr}).  When rotation powered,
the three known tMSPs are redbacks.  Searches of yet unidentified {\it
  Fermi} gamma-ray sources \cite{salv17,str19} with suitable spectral parameters have turned out to be the main
technique to discover these otherwise elusive eclipsing radio MSPs.

\section{Changes of state in millisecond pulsars}
\label{sec:changes}

The multifaceted behaviour of MSPs in tight binaries stems from the
balance between the outward pressure exerted by the pulsar wind on the
mass lost by the companion star, and the inward gravitational pull
applied by the NS gravitational field. Given the typical MSP spin-down
power ($L_{sd}\simeq \mbox{a few} \times 10^{34}\,\ergsec$) and mass
transfer rates ($\dot{M}\approx 10^{-3}-10^{-4}\,\dot{M}_{Edd}$), this
balance enforces within the binary, if the system has a short orbital period
($P_{orb}<1$~day) and a small size ($d\simlt 10^{11}$~cm). On one hand,
this means that the pulsar wind is terminated by the in-flowing  matter
in an intrabinary shock. On the other, slight variations in the mass
inflow rate may lead to very different outcomes, which in the case of
tMSPs  occur in quick succession.

\begin{figure}[t!]
\includegraphics[scale=0.56]{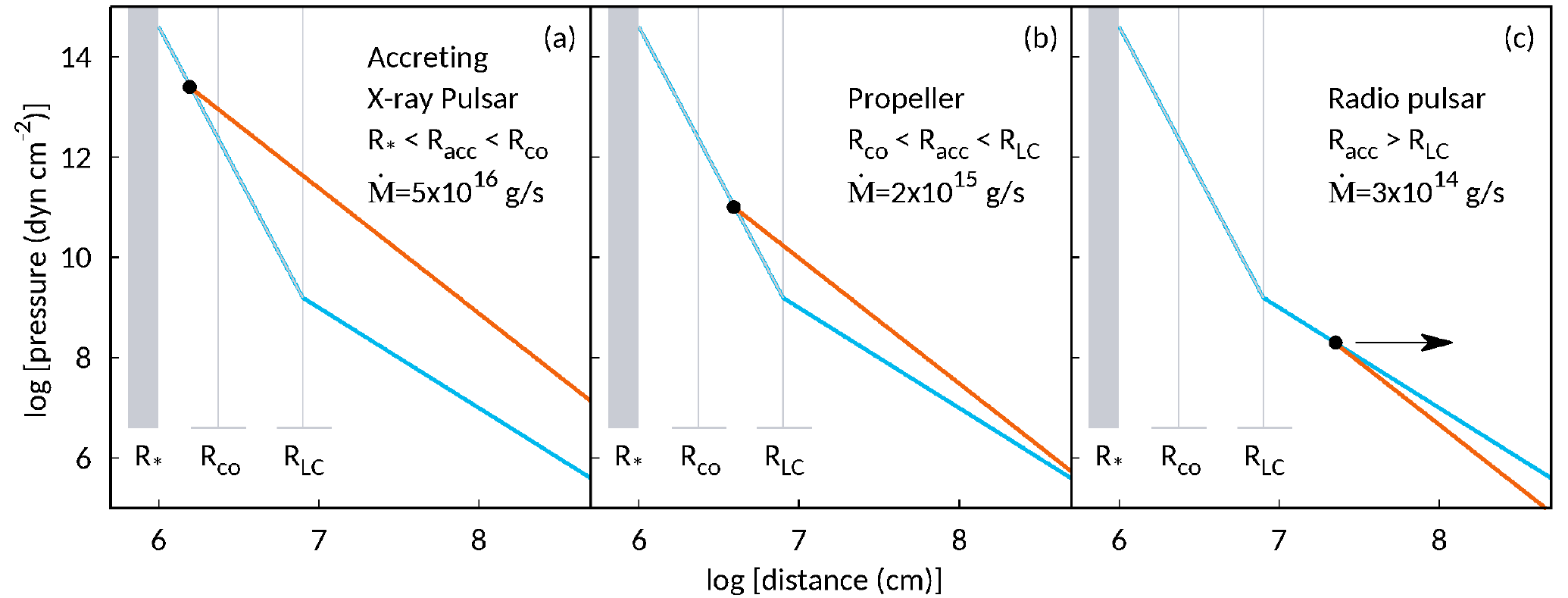}
\caption{Dependence of the pressure exerted by the in-flowing plasma
  (orange lines, Eq.~\ref{eq:pmat}) and by the electromagnetic field
  (cyan lines, Eq.~\ref{eq:pem1} and \ref{eq:pem2}) on the distance
  from a pulsar with a magnetic dipole moment of $10^{26}$~G~cm$^3$
  (corresponding to a field strength at the magnetic poles of $5\times10^7$~G). 
  Grey vertical lines mark the NS radius ($R_*$),
  the co-rotation radius ($R_{co}$) and the light cylinder radius
  ($R_{LC}$) for a pulsar spinning at 600~Hz (corresponding to $\simeq
  1.67$~ms). The three panels represent the possible states depending
  on the value of the mass accretion rate considered, accreting X-ray
  pulsar ($R_{*}<R_{acc}<R_{co}$), propeller ($R_{co}<R_{acc}<R_{LC}$)
  and rotation-powered radio pulsar ($R_{acc}>R_{LC}$). The arrow
  indicates evaporation of the intrabinary matter due to the unstable
  equilibrium in the latter case.}
\label{fig:pressure}       
\end{figure}

In the recycling framework, it was assumed that the accretion and the
rotation-powered phases were well distinct. A source was to be found
in either one of the states depending on the prevailing of the
gravitational or electromagnetic pressure
\cite{shv70,lip87}. Fig.~\ref{fig:pressure} shows the radial
dependence of these pressures and the three main possible outcomes. The
ram pressure of plasma in radial free-fall is \cite{dav73}:
\begin{equation}
P_{grav}=\frac{(2GM_*)^{1/2}\dot{M}}{4\pi r^{5/2}},
\label{eq:pmat}
\end{equation}
where $M_*$ is the NS mass. In the accretion phase, the high-density
plasma fills the light cylinder of the pulsar\footnote{The light
  cylinder of a pulsar is defined as the cylinder aligned with spin
  axis, with a radius $R_{LC}=c/2\pi\nu =c/2\pi\nu\simeq
  80\,\nu_{600}$~km ($\nu_{600}$ is the spin frequency in units of
  600~Hz), where the co-rotating speed of the field lines equals the
  speed of light. Field lines that would close beyond the light
  cylinder are forced to open by the causality principle, thus
  producing the pulsar wind.}  and this was assumed to switch off the
rotation-powered pulsar \cite{shv71}. Some of the closed field lines
forming the magnetosphere thread the disc albeit the very
high diffusivity of the plasma, and are bent by the differential
rotation of the disc material in Keplerian rotation. The disc is
truncated at the accretion radius $R_{acc}$, where the resulting magnetic stress becomes dominant
compared to the disc viscous stress. The determination of this radius is crucial to predict the accretion regime onto
a magnetised rotator at different accretion rates. It strongly
depends on the often unknown micro-physics governing the disc/field
interaction and is still a matter of debate (see, e.g.,
\cite{boz09,boz18}). Under the assumption that the magnetic field lines  thread the disc over a large radial extent \cite{gho79,wan95}, it
turns out that the accretion radius $R_{acc}$  is equal to a fraction $\xi$ of the Alfven radius $R_A$, obtained by equating the gravitational energy density (see Eq.~\ref{eq:pmat}) with the largest possible magnetic stress:
\begin{equation}
  P_{em}(r)=\frac{\mu^2}{8\pi r^6};\;\;\;\; r<R_{LC}
  \label{eq:pem1}
\end{equation}
where $\mu=B_* R_*^3 /2$ is the NS magnetic dipole moment, $B_*$ is the
field strength at the magnetic poles of the NS, and $R_*$ is the NS
radius. This yields:
\begin{equation}
  R_{acc} = \xi R_{A} = \xi \frac{\mu^{4/7}}{\dot{M}^{2/7}(2GM_*)^{1/7}} \simeq 15.4\, \xi_{0.5}\, \mu_{26}^{4/7} \,\dot{m}_{16}^{-2/7}\,m_{1.4}^ {-1/7}\,\mbox{km}, 
  \label{eq:racc}
\end{equation} 
where $\xi_{0.5}$, ${\mu}_{26}$, $\dot{m}_{16}$ and $m_{1.4}$ are the
respective quantities in units of 0.5, $10^{26}$~G~cm$^3$,
$10^{16}$~g~s$^{-1}$ and 1.4~$\msun$. Alternatively, if the
diffusivity of the field lines is low, the magnetic field lines are unable
to slip through the disc fast enough, and so they  twist and open. The
interaction region is smaller and the scaling of the accretion radius
are flatter, $R_{acc}\propto \mu^{2/5} \dot{M}^{-1/5}$
\cite{spr93,dan10}. A series of 3D magneto-hydrodynamics simulations of
disc accretion onto a magnetised rotating star have indeed given
results compatible with such an estimate \cite{kul13}. However, given
the small differences, Eq.~\ref{eq:racc} with $\xi\simeq0.5$ is still
widely used to describe the position of the magnetospheric radius in
many astrophysical systems (see, however, \cite{boz09} for the limitations of the applicability of the magnetospheric radius in Eq.~\ref{eq:racc}).

The subsequent fate of the in-flowing matter depends on whether it has enough
angular momentum to overcome the centrifugal barrier set by the
rotating pulsar magnetosphere at the disc truncation radius. The
co-rotation radius defines where the disc material rotates at the NS
spin frequency:
\begin{equation}
  R_{co}=\left[\frac{GM_*}{(2\pi\nu)^2}\right]^{1/3}\simeq 23.6 \,m_{1.4}^{1/3}\,\nu_{600}^{-2/3}~{\rm km}.
    \label{eq:rco}
\end{equation}
 The in-flowing material  either freely accretes onto the NS if
 $R_{acc}<R_{co}$ (see panel [a] in Fig.~\ref{fig:pressure}) or
 bounces on the barrier set by the rotating magnetosphere in the
 so-called propeller state if $R_{acc}>R_{co}$ \cite{ill75} (see panel
 [b] in Fig.~\ref{fig:pressure}).
 
As the mass accretion rate decreases, the accretion radius
expands. When it approaches the co-rotation boundary, no net angular
momentum can be transferred to the NS by the in-falling plasma any more. Nevertheless,  the field lines which  thread the disc beyond the co-rotation
radius, and possibly eject matter through the propeller effect, might still exert a spin-down torque
\cite{gho79,wan95,rap04}. This torque is assumed to limit the secular spin-up
of an accreting NS in an LMXB to an equilibrium period of a few
milliseconds for a weakly magnetised ($\approx10^8-10^9$~G ) NS.  Realising that the equilibrium period
could be so short was actually the main foundation of the recycling scenario
\cite{sri82}.

A further decrease of the mass accretion rate may push the accretion
radius beyond the light cylinder, at which point a rotation-powered
pulsar is assumed to switch on (see panel (c) in
Fig.~\ref{fig:pressure}). Outside the light cylinder, a radiative
solution describes the pulsar electromagnetic field and its pressure
has a much flatter radial dependence than Eq.~\ref{eq:pem1}:
\begin{equation}
P_{em}(r) =\frac{L_{sd}}{4\pi r^2 c} \simeq \frac{k \mu^2}{4\pi^2 R_{LC}^4 r^2};\;\;\;\; r>R_{LC}.
\label{eq:pem2}
\end{equation}
Here $L_{sd}\simeq [\mu^2 (2\pi \nu)^4 / c^3] (1+\sin^2\alpha)$ is the
pulsar spin-down power \cite{spi06}, and $\alpha$ is the magnetic
co-latitude. Since the inflow ram pressure has a steeper dependence on
the distance ($P_{grav}\propto r^{-5/2}$, see Eq.~\ref{eq:pmat}), the
equilibrium for $R_{acc}>R_{LC}$ is expected to be unstable. In this regime the pulsar wind is then expected to expel altogether the intrabinary material from the
system and switch-off the mass transfer.  In the secular picture, this is expected to take place when the  mass transfer rate drops as the donor star detaches from its Roche lobe under the irradiation by the high energy
emission of the accreting NS \cite{klu88,rud89,rud89b}.

The idea that binary systems could perform state transitions between
accretion and rotation-powered states on a shorter timescale of
months/years emerged when many LMXBs hosting NSs were found to undergo
transient outbursts (see, e.g., \cite{whi89}). At the end of an
outburst, the pulsar wind was expected to push the accretion disc
beyond the light cylinder (i.e., case (c) of
Fig.~\ref{fig:pressure}). This would allow the turn-on of a rotation-powered radio
pulsar during the quiescent period of the transient \cite{ste94,cam98}. The radiation and the 
wind of relativistic particle so generated would eject the matter from
the companion as soon as it entered the pulsar Roche lobe - the
so-called radio ejection mechanism \cite{bur01}. Only an increase of
the inward pressure of the transferred matter would allow the X-ray
binary to enter in a new accretion outburst.  For many years, only
circumstantial evidence supported this idea. A few eclipsing MSP
binaries were found to expel material transferred from their
companions \cite{dam01} according to a radio ejection scenario
\cite{bur02}, but no transition to an accretion stage was
observed. On the other hand, accreting MSPs in quiescence gave
indirect indications that a radio pulsar switched on, but a detection
could not be achieved (see  Sec.~\ref{sec:outburst}).  The
discovery of tMSPs has eventually filled the gap.

\section{Transitional millisecond pulsars}
\label{sec:sources}

\begin{table*}
\flushleft
 \begin{minipage}{140mm}
  \caption{\label{tmsp_prop} Main system properties of tMSPs}
  \begin{tabular}{@{}lcccc@{}}

  & PSR\,J1023+0038  & XSS\,J1227-4859 & IGR\,J1824-2453 & Ref.\\
\hline
      &   & & & \\
 $P_{orb}$ (h)         & 4.75  & 6.91  & 11.03 & \cite{tho05,bas14,pap13}\\
 $P_{spin}$ (ms)       & 1.69  & 1.69  & 3.93   &\cite{arc09,roy15,pap13}\\
$\dot{P}$ (10$^{-20}$) &  0.539 (RMSP) - 0.713 (LMXB)$^A$  & 1.086 & $<0.0013$ & \cite{arc13,jao16,roy15,pap13}\\ 
 $a\,sin i$ (lt-s)      & 0.343 & 0.668 & 0.766 &\cite{arc09,jao16,pap13}\\
$\dot{E}$$^B$ ($\rm 10^{34}\,erg\,s^{-1}$) & 4.43(4) & 8.9$^{+0.2}_{-0.9}$ & - & \cite{arc13,dem20}\\  
$ M_{NS}$ ($\rm M_{\odot}$) & 1.7(2)  & -      & - & \cite{sha19}\\ 
$ B_{NS}$ ($10^8$\,G)  & 1.9   & 2.3 & 0.7-35 & \cite{arc13,roy15,pap13} \\
$i$ (deg)                 & 46(2) & 46-55 & -  & \cite{sha19,dem14,riv18,str19}\\
d (kpc)                   & 1.37(4) & 1.4$^{+0.7}_{-0.2}$ & 5.5 & \cite{del12,dem20,pap13}\\
Companion spectral type   & G5$->$F6 & G5$->$F5 & low main sequence & \cite{tho05,dem14,pal13}\\ 
Companion mass ($\rm M_{\odot}$)        & 0.22(3)  & 0.15-0.36 & 0.17$^C$ & \cite{sha19,dem15,pap13}\\

& & &  \\
{\it X-ray properties}\,$^D$   $\rm F_{E} \sim E^{-\Gamma}$ & &  & & \\ \hline
{\it Disc state} & &  & &\\
$\rm \Gamma_{X,disc}$   & 1.62(2) & 1.70(2) & 1.428(3) & \\
${L_X,ave}$ ($\rm 10^{33}\,erg\,s^{-1}$) & 5.2(1) & 12(2) & 11.2(5) & \\
${L_X,low}$ ($\rm 10^{33}\,erg\,s^{-1}$) & 0.87(4) & 2.0(4) & 2.0(3) & \\  
$ {L_X,high}$ ($\rm 10^{33}\,erg\,s^{-1}$)& 7.9(1) & 13(3) & 13.1(6) & \\ 
$ {L_X,flare}$ ($\rm 10^{33}\,erg\,s^{-1}$) & 22(6) & 70(8) &  33 & \\ 
{\it Rotation-powered state} & &  & \\ 
$ \Gamma_{X,rot}$ & 1.17(9) & 1.2(1) & 2.5 assumed & \\
${L_{X,rot}}$ ($\rm 10^{33}\,erg\,s^{-1}$) & 0.8(4)  & 0.83$^{+0.42}_{-0.9}$ & $< 1$ & \\
& &  & & \\ 

{\it Gamma-ray properties}\,$^E$  $\rm F_{E} \sim E^{-\Gamma}$& &  & & \\ \hline
{\it Disc state} & &  & & \\ 
$\rm \Gamma_{\gamma,disc}$  & 2.41(2)(3) & 2.36(6)(9) & - & \\ 
$ {L_\gamma,disc}$ ($\rm 10^{33}\,erg\,s^{-1}$) & 12.5(4) & 21.9(7) &  - & \\
{\it Rotation-powered state} & &  & & \\
$\rm \Gamma_{\gamma,rot}$ &  2.31(3)(4) & 2.42(3)(15) & - & \\
${L_\gamma,rot}$ ($\rm 10^{33}\,erg\,s^{-1}$) & 1.1(2) & 8.6(8) &  - & \\
& &  & & \\

{\it Radio properties}\,$^F$ $\rm S_{\nu} \sim \nu^{\alpha}$ & &  & & \\ \hline
{\it Disc state }  & &  & & \\
$\alpha_{\rm r,disc}$ & -0.1(2)$->$0.2(2) & -0.1(0.1) & -0.2$->$0.8 & \\ 
${L_r,disc,ave}$ ($\rm 10^{27}\,erg\,s^{-1}$) & 0.97 & 2.4(8) & 123 & \\ 
$ {L_r,disc,high}$ ($\rm 10^{27}\,erg\,s^{-1}$) & 0.63 & -   &  - & \\
$ {L_r,disc,low}$ ($\rm 10^{27}\,erg\,s^{-1}$) & 1.9  & - & 244 & \\
{\it Rotation-powered state} & &  & & \\
$\rm \alpha_{r,rot}$ & 2.8 & --  &  -- & \\
${L_r,rot,ave}$ ($\rm 10^{27}\,erg\,s^{-1}$) & 3800 & $<$0.37 & - & \\
\hline
\end{tabular}

$^A$: Spin down rates in both RMSP and LMXB state determined only for  {\psr1023}, taking into account also the Shklovskii effect and acceleration in the Galactic potential. 
$^B$ Spin down luminosity assuming the canonical value for the moment of inertia $\rm 10^{45}\,g\,cm^2$ \footnote{The NS moment of inertia has been found to range from 1-4$\rm \times 10^{45}\,g\,cm^2$ by detailed general relativistic numerical computations using a sample of MSPs with precise NS masses and for several realistic NS EoS models \cite{bat17}. }
 $^C$: Minimum mass.
 $^D$: X-ray luminosity in the 0.3-79\,keV range from \cite{cze19}. 
$^E$: Gamma-ray luminosity in the 0.1–300\,GeV using fluxes 
from \cite{tor17} and the reported distances at face value.
$^F$: Radio luminosity at 5\,GHz using distances reported at face value and fluxes from \cite{pap13,fer14} for IGR\,J1824-2452, 
from \cite{arc09,arc13,bog18,pap19,cze19} for
{\psr1023}, from \cite{bas14,roy15} for PSR\,J1227-4853. 
\end{minipage}
\end{table*}

\subsection{{\psr1023} -- FIRST J102347.6-003841 }

FIRST J102347.6-003841 was first detected in May 2000 as a variable
1.4~GHz radio source \cite{bon02}. Double peaked emission lines of the
Balmer series, He I and He II in the spectrum and the flickering light
curve of its optical counterpart suggested it was a disc accreting
binary, possibly a peculiar magnetised white dwarf
\cite{bon02,szk03,wan09}.  In early 2002, the source underwent a
dramatic change. An almost sinusoidal smooth modulation at the 4.75~hr
orbital period due to heating of the secondary appeared in the
optical light curve \cite{wou04}. Also, emission lines were replaced by
a G-type absorption spectrum \cite{tho05,wan09} indicating the
disappearance of the disc. Thorstensen et al. \cite{tho05} first
proposed that the binary hosted a quiescent NS, based on the large
irradiating luminosity required to explain the optical light curve.
The discovery of the 1.67~ms radio pulsar {\psr1023} by the {\it
  Robert C. Byrd Green Bank Telescope} in 2007 eventually nailed down
this enigmatic object as a redback radio MSP, which had an accretion
disc in the previous decade \cite{arc09}. Archival optical and
infrared observations constrained the duration of the 2000/2001 disc
episode to $\sim1.5-2$ years \cite{wan13,bon02,szk03,wou04}. Starting
at the end of June 2013, the re-emergence of double-peaked optical
emission lines \cite{hal2013}, the disappearance of radio pulsations
\cite{sta14}, and the brightening of the X-ray, ultraviolet \cite{pat14} and gamma-ray \cite{sta14} emissions
marked the beginning of a new active phase
which is currently ongoing in June 2020.  The average X-ray luminosity
in the disc state never exceeded $L_X\approx 5 \times 10^{33}\,\ergsec$ (at a
parallax distance of 1.37~kpc, \cite{del12}), indicating that both the
2000/2001 and the 2013/current accretion episodes have been {\it
  sub-luminous} (see Sec.~\ref{sec:sublum}).

\subsection{IGR J18245-2452 -- PSR J1824-2452I}

The transient X-ray source IGR J18245--2452 in the globular cluster
M28 was first detected by {\it INTEGRAL} in March 2013 during a bright
 accretion
outburst (\cite{eck13}, $L_X\approx10^{36}\,\ergsec$ at a distance of 5.5~kpc). The {\it
  XMM-Newton} detection of 3.9~ms X-ray pulsations identified it as an
accreting MSP with a main sequence companion
\cite{pap13}. Cross-referencing with pulsar catalogues, it was realised
that the source had been already observed as a radio MSP before (PSR
J1824-2452I; \cite{beg06}), making it the first source both as a
rotation-powered and as an AMXP  \cite{pap13}. Radio pulses
were again observed after the end of the month-long X-ray outburst in
2013 -- two weeks since the last detection of the X-ray
pulsar. Serendipitous {\it Chandra} \cite{pap13,lin14} and {\it Hubble
  Space Telescope} observations \cite{pal13} revealed two more
accretion episodes in 2008 and 2009, respectively, which unlike the
2013 outburst, had properties compatible with a {\it sub-luminous}
disc state.

\subsection{{\xss} -- PSR J1227--4853}

{\xss} resembles very closely {\psr1023} under many respects (see
Table ~\ref{tmsp_prop}).  First detected as a hard X-ray source
\cite{saz04}, it was tentatively identified as a cataclysmic variable
based on the emission lines of its optical spectrum \cite{mas06} and
for large amplitude ($\sim 1$~mag) optical flickering
\cite{pre09}. The spatial coincidence of {\xss} with a {\em Fermi}-LAT
source suggested an atypical low-luminosity ($L_X\simeq
L_{\gamma}\simeq\mbox{few}\times10^{33}\,\ergsec$) X-ray binary
\cite{dem10,dem13}. Its unusual properties were only later assessed to
be typical of the {\it sub-luminous} state of tMSPs  (\cite{sai09,dem10,hil11,dem13,dem14}, see
Sec.~\ref{sec:states}).  The
disappearance of the emission lines in the optical spectrum and the
10-fold dimming observed in the radio, optical and X-ray bands (and to
a lesser extent in gamma-rays, \cite{tor17}) demonstrated that
{\xss} had transitioned from a disc to a radio pulsar state between
2012 November 14 and December 21 \cite{bas14}. {\it Giant Metrewave
  Radio Telescope} observations later detected 1.69~ms radio
pulsations eclipsed for a large fraction of the orbit \cite{roy15}.
Since the end of 2012, {\xss} still behaves as a rotation-powered
redback pulsar.

\section{The three states of transitional millisecond pulsars}
\label{sec:states}

\subsection{The rotation-powered state}
\label{sec:radiopsr}

In the rotation-powered state, tMSPs behave as redbacks.
They are relatively faint objects at all
wavelengths, and most of the information we have gathered comes  from the study of the  closest ones, {\psr1023} and PSR\,J1227--4853 (see Table\,1).

{\bf Mass ejection --} Irregular eclipses of the radio pulses
occur mostly (but not only) when the secondary is at the inferior
conjunction of the orbit. They are due to a thin, but dense layer of ionised
material which the pulsar wind drives off from the surface of the donor or  the
inner Lagrangian point, and partly enshrouds the system. The eclipses of {\psr1023} lasted up to $\sim60\%$ of the orbit
when observed at 350~MHz, but were shorter at higher frequencies
($\sim25\%$ at 1.4~GHz), and nearly absent at $\sim3$~GHz
\cite{arc09,arc13}. Similarly behaved PSR\,J1227--4853, showing
eclipses for $\sim 40\%$ of the orbit at 607\,MHz and $\sim30\%$ at
1.4~GHz \cite{roy15,dem20}. This can be ascribed to the frequency
dependence of the optical depth of the material (e.g., $\propto
\nu^{-2}$ for electron scattering, $\propto \nu^{-1}$ for cyclotron
absorption \cite{tho94}) which makes the ionised layer more
transparent at high frequencies. The correlated variation of the
continuum flux density and the mean pulsed flux density at the eclipse
boundary observed from PSR\,J1227--4853 suggested
cyclotron-synchrotron absorption rather than dispersion smearing or
scattering \cite{roy15}.  Similar conclusions were also drawn for
other spiders \cite{bro16}.
Shorter losses of the signal at random orbital phases and substantial variations of the dispersion measure were also observed.  All
these properties indicate that the enshrouding ionised plasma extends
well beyond the Roche lobe of the donor.

 {\bf The binary properties} -- The radio pulsar timing indicated that the
 orbits of tMSPs and of many spiders are almost circular, with upper
 limits  of the order of a few times $10^{-5}$ on the eccentricity.
 This is due to the tidal circularization which occurred during the secular
 LMXB phase. Irregular changes in the phase of the orbital
 modulation by a few seconds over timescales of a few months suggested fast apparent orbital period variations
 \cite{arc13,jao16}. They have been interpreted as a
 combination of the angular momentum carried by material ejected 
 from the system, and the exchange of angular momentum between the
 orbit and the companion star due to changes of the mass quadrupole
 in the latter \cite{app94}. Magnetic cycles of the secondary
 could be the cause for such fluctuations, even though luminosity
 variations larger than observed would be expected.
Recently, a model has been proposed to account for the timing
anomalies produced by mass quadrupole deformations in spiders, 
which is promising to gain insights on the internal structure of their
irradiated stars  \cite{voi20}.

The optical spectra of {\psr1023} and PSR\,J1227--4853 
displayed  absorption
features typical of the
photosphere of mid G-type stars, primarily metallic and Balmer lines \cite{tho05,dem14,mcc15,sha19}.
No spectroscopy has been acquired for the faint 
counterpart of PSR\,J1824-2452I, so far. However, photometric observations located it
in a position of the colour-magnitude diagram of the globular cluster
consistent with a main-sequence star, just 0.5-1\,mag below the
turn-off point \cite{pal13}. The lack of a an optical and near-infrared
polarisation in PSR\,J1227--4853
suggested that the emission in these bands
was only due to the donor photosphere
\cite{bag16}. However, near-ultraviolet photometry of both {\psr1023}
and PSR\,J1227--4853 revealed an excess  over the companion
emission \cite{riv18}. This suggests that the emission of the intrabinary shock, which dominates at
higher energies (see below), may also extend to the ultraviolet
domain.

\begin{figure}[t!]
\includegraphics[scale=1.6]{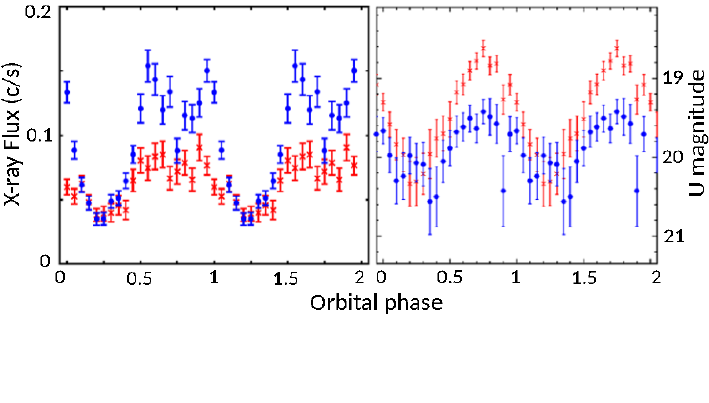}
\caption{X-ray (left panel) and optical (right panel) orbital
  modulation observed from PSR\,J1227--4853 on 2013, Dec 29 (red
  points) and 2014, Jun 27 (blue points). Phase 0 corresponds to the   passage of the NS at the orbit ascending node. Adapted from \cite{dem15}.}  \label{fig:radiopulsarmod} 
\end{figure}

The optical light curves of these two tMSPs
featured an almost sinusoidal modulation at the binary orbital period
with an amplitude of $\approx 0.4-0.7$~mag. The emission attains a the maximum and becomes bluer when the companion star is at the superior conjunction (\cite{wou04,tho05,dem15,riv18}; see the right panel of
Fig.~\ref{fig:radiopulsarmod}). These signatures are due to the  
irradiation of the donor by the high-energy emission of the pulsar and were also observed from many spiders \cite{bre13,rom16}. A dramatic change in the spectral
type (see Table\,1) between the inferior and the superior conjunction confirmed the heating of the donor by the wind of tMSP 
\cite{dem14,sha19}.  Variations in the amplitude of the orbital
modulation from epoch to epoch also suggested changes in the heating
pattern \cite{dem15,sha19}. In both systems the irradiation  persisted also during the
disc state  \cite{dem14,cze14,bog15,pap18,ken18} with
a different heating pattern likely due to disc shadowing of the donor
star \cite{sha19}.

 Modelling of the  multi-colour optical orbital modulation  and of the spectroscopic absorption line radial velocity is a  technique commonly used to infer the component
 masses, the Roche lobe filling factor of the companion, and the binary
 inclination. It is especially powerful when coupled with the NS orbital ephemeris
 derived from the radio pulsar timing.  However, for heated low-mass
 donors a degeneracy exists between the filling factor and the binary
 inclination, thus affecting the masses of the components \cite{str19}. The
 two tMSPs in the Galactic field are seen at moderate inclinations
 \cite{tho05,dem15} (see Table\,1). While in {\psr1023} the masses of  both
 stars could be determined, with an indication of an
 underfilling donor \cite{mcc15,sha19}, for PSR\,J1227-4853 only
 the donor mass could be constrained.

{\bf The X-ray properties and the intrabinary shock --} The X-ray luminosity  of {\psr1023} \cite{hom06,arc10,bog11,ten14, li14}, PSR
J1227--4853 \cite{bog14,dem15,dem20} and PSR M28I
\cite{bog11a,pap13,lin14} in the radio-pulsar state ($L_X\approx 1-2\times10^{32}$ \ergsec, i.e. $\sim 0.1-0.2\%$ of
the spin-down power) is at the bright end of the distribution observed
in other redbacks (\cite{lin14b}, see the bottom panel of
Fig.~\ref{fig:lin14b}).  The X-ray spectra of tMSPs were largely non-thermal
and described by a power law extending without a break
up to $\sim$70~keV (see Table\,1). A soft thermal component possibly emitted by hot
spots on the NS surface contributed at most to a few per
cent of the 0.1--10 keV emission of {\psr1023}, while it was not
significantly detected in PSR\,J1227-4853.

A large amplitude ($\simgt
25\%$) orbital variability characterised the X-ray emission of tMSPs 
(see the left panel of Fig.~\ref{fig:radiopulsarmod}), similar to other redbacks \cite{rob18}. The maximum
occurs when the companion is at the superior conjunction, in phase
with the radio eclipses and the maximum of the optical emission. Only
a slight spectral variability along the orbit has been seen, e.g. a
hardening below 3\,keV at the inferior conjunction of {\xss}, and a
decrease in the amplitude above 25~keV (\cite{dem20}, Coti Zelati et
al., in prep.). This excluded photoelectric absorption to explain 
the large orbital variability. The observed emission was better modelled with a synchrotron
emission from the intrabinary shock created by the interaction of the
pulsar wind with the material which issues from the inner Lagrangian point, or
directly at the donor surface (\cite{aro93,bog11}, see
Sec.~\ref{sec:radiopulsarmodels}). The occultation of the shock by the
secondary star when it is at the inferior conjunction of the orbit determines the minimum of the X-ray emission.  The orbital
modulation was almost sinusoidal in {\psr1023} with an enhanced
emission at eclipse egress \cite{arc10,bog11}. In PSR J1227--4853,
instead, the orbital light curve showed a quasi-sinusoidal shape at
one epoch but double-peaked when observed 6\,months apart. Concurrently, the amplitude of the X-ray modulation varied from  from 25$\%$
to 70$\%$, in anti-correlation with a similar change in the amplitude of the  optical modulation (\cite{dem15}, see
Fig.~\ref{fig:radiopulsarmod}). Double peaked orbital modulations were
also observed from other redbacks such as PSR\,J2129-0429 \cite{aln18}
and PSR\,J2339-0533 \cite{kan19}, which also displayed subtle spectral
changes along their orbits. The synchrotron emission is expected to be
Doppler boosted at the pulsar inferior conjunction and de-boosted
at superior conjunction, possibly explaining  these features, at least partly (\cite{aro93,dub15}, see also
Sec.\ref{sec:radiopulsarmodels}).

X-ray pulsations with a sinusoidal shape and a root-mean-square
amplitude of $(11\pm2) \%$ were detected below 2.5~keV from {\psr1023}
\cite{arc10}.  The pulsed luminosity was a few $\times10^{-4}$ times
the spin-down power, similar to other rotation-powered X-ray MSPs
\cite{pos02}.  Sinusoidal profiles are usually ascribed to the heated polar caps on the NS surface \cite{zav07}, although the thermal component observed in the
X-ray spectrum was too faint to account for the observed pulse
amplitude. A pulsed signal was not detected instead from either PSR J1227--4853
(within an upper limit of 10\%, \cite{pap15}) or PSR M28I
(for which high time resolution data lacked \cite{pap13,lin14}).

{\bf The gamma-ray emission and particle acceleration --} MSPs are
energetic enough to convert a few per cent of their spin-down power
into emission at GeV energies, and TMSPs made no exception \cite{tor20}. They were characterised by a luminosity
of a few $\times\,10^{33}\ergsec$ (0.1--100~GeV) and a power law
spectrum with photon index $\sim 2.3-2.4$
(\cite{tam10,nol12,tak14,tor17,joh15}, see Table\,1).  A marginally
significant high energy cut-off at $\sim 5$~GeV could be detected only
in PSR\,J1227--4853 \cite{tor17}. Gamma-ray pulsations were reported
at a significance of 3.7$\sigma$ from {\psr1023} \cite{arc13} and
5$\sigma$ from PSR J1227--4853 \cite{joh15}. The gamma-ray
($>$100 MeV) pulse profile featured a relatively broad peak almost aligned
with the main peak of the radio pulse at 1.4\,GHz. While an orbital
modulation was not detected during the rotation-powered state of
{\psr1023}, its presence in PSR J1227--4853 is controversial
\cite{xin15,joh15}.

\subsection{Accretion outbursts}
\label{sec:outburst}

\begin{figure}[t!]
\includegraphics[width=1.0\textwidth]{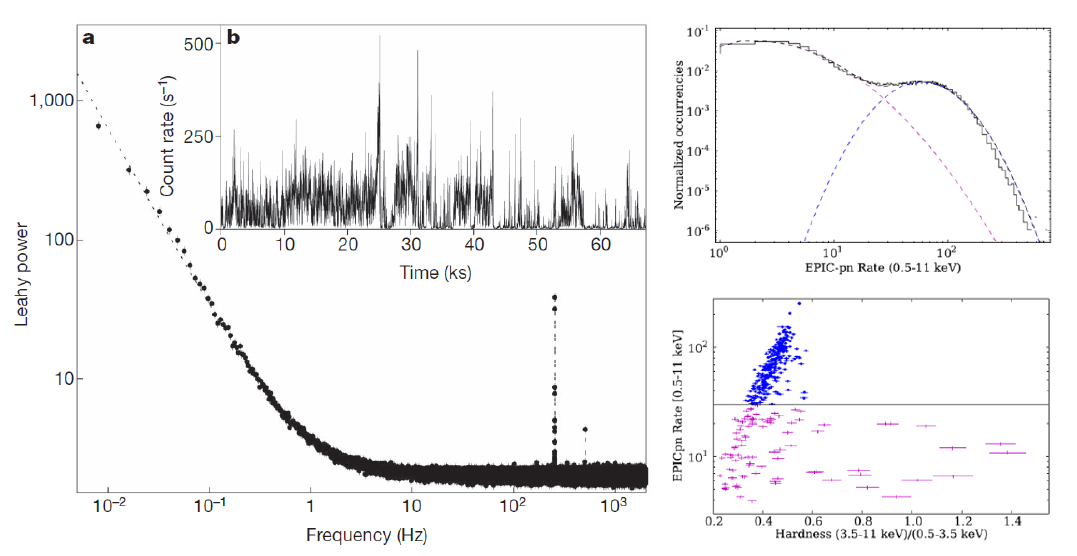}
\caption{Left panel: X-ray power density spectrum and light curve 
  (inset) of {\igr} observed by {\it XMM-Newton} during its 2013
  outburst. Right panel: Histogram of the X-ray count rates (top
  panel) and hardness-intensity diagram (bottom panel) showing the two
  states. Figures are taken from \cite{pap13,fer14}.}\label{fig:m28i}
\end{figure}

So far, {\igr} has been the only tMSP that has shown an accretion
outburst with a similar peak X-ray luminosity ($L_X\simeq 5 \times
10^{36}\,\ergsec$) and duration ($\sim$ three weeks) than other AMXPs. However, a peculiar and extremely strong variability was seen in two
$\sim$ one day-long {\it XMM-Newton} observations performed a few days
apart  (\cite{pap13}, see the left panel of Fig.~\ref{fig:m28i}). It had an RMS amplitude of more than 90\% and its Fourier power density spectrum was described with a
power law $P(\nu)\propto \nu^{-\gamma}$ with index $\gamma=1.2$,
extending over six decades in frequency ($10^{-4}-100$~Hz).
Two states could be identified with a flux differing by two
orders of magnitude (\cite{fer14}, see the right panels of
Fig.~\ref{fig:m28i}).  In the high-intensity state ($L_X\simeq \mbox{a
  few} \times 10^{36}\ergsec$, corresponding to a mass accretion rate
of $\sim 10^{16}$~g~s$^{-1}\simeq 10^{-2}\,\dot{M}_{Edd}$), a power law
$F(E)\propto E^{-\Gamma}$  with $\Gamma\simeq 1.7$
characterised the X-ray energy spectrum. This is typical of AMXPs and
is usually interpreted as Compton up-scattering of soft photons coming
from the NS surface by hot electrons in the accretion columns. The spectrum became significantly
harder, $\Gamma\simeq 0.9$, in the low-intensity state ($L_X\simeq \mbox{a
  few} \times 10^{34}\ergsec$, corresponding to $\dot{M}\sim
10^{14}$~g~s$^{-1}=10^{-4}\,\dot{M}_{Edd}$), and an 
additional partial covering absorption component was required. Together with
a strongly variable emission observed at GHz radio frequencies, this
suggested the presence of out-flowing material \cite{fer14}. The spectral hardening characterising the low-intensity state also made the average spectrum
of {\igr}  the hardest of all AMXPs \cite{def17}. The X-ray
pulse profile showed two sinusoidal peaks per cycle.  Pulsations were
present at all flux levels up to 60~keV, and the amplitude correlated
with  flux up to an rms of $\sim 20\%$. In the lower
intensity-harder flux state the amplitude was instead lower ($\sim
5\%$) and the shape different. This peculiar behaviour was interpreted
in terms of a fast switching between accretion and weak to strong
propeller states \cite{rom04,fer14}.

{\bf Other accreting MSPs in quiescence --} {\igr} has been the only
AMXP detected as a radio pulsar in quiescence, so
far. However, many indirect pieces of evidence suggest that a radio pulsar may
have switched on also in other AMXPs. When in quiescence, they are
relatively faint compared to other soft X-ray transients
($10^{31}-10^{33}\,\ergsec$; see, e.g., \cite{wij17}) and show a
non-thermal power-law spectrum, possibly originated from an
intrabinary shock \cite{cam02}. On the other hand, a soft thermal component
was found to be hardly detectable, giving stringent constraints on the
rapidity of the cooling of the NS atmosphere after an accretion
outburst \cite{hei09}. Such a faint X-ray emission could not account
for the optical flux of the irradiated companion, requiring the more intense
spin-down power of a rotation-powered pulsar
\cite{bur03,cam04,dav09}. AMXPs in quiescence spin down at a rate
compatible with magneto-rotational torques \cite{har08} and their
orbit showed a rapid and complex evolution similar to black widow
pulsars, possibly due to the ejection of matter from the system
\cite{dis08} and/or angular momentum exchange between the binary and
the donor \cite{pat17}. {\em Fermi}-LAT data also unveiled a gamma-ray
counterpart of the closest ($d=3.5$~kpc) AMXP known,
SAX~J1808.4--3658 \cite{deo16}. The luminosity measured during the ten years between August 2008 and 2018 (in which three outbursts have also occurred) was $L_{\gamma}=(6\pm1)\times10^{33}\,\ergsec$. This is compatible with the values observed from rotation-powered MSPs, although 
pulsations could not be detected.

However, despite thorough searches, radio pulses were not detected
from AMXPs other than {\igr} \cite{burg03,iac09,iac10}, down to an
upper limit of $30\,\mu$Jy (at 2~GHz) in the case of SAX~J1808.4--3658
\cite{pat17}. An unfavourable inclination (although the radio beams of
MSPs are very large), absorption of radio waves at low frequencies
(where most of the radio power is emitted) by matter enshrouding the
binary, and/or the larger distances of AMXPs (most are located in the
Galactic bulge) than transitional systems are possible reasons.  Worth
noticing is that {\igr}  was sporadically detected as a faint radio
pulsar with a flux density of $10-20\,\mu$Jy at 2~GHz, i.e. close to
the sensitivity limit, because the M28 cluster where it resides is rich
of MSPs, and thus was deeply surveyed in the radio domain
\cite{beg06,pap13}.

\subsection{The sub-luminous disc state}
\label{sec:sublum}

All the three tMSPs  showed an enigmatic accretion disc state
characterised by an X-ray luminosity of $\sim 10^{33}-10^{34}\,\ergsec$, fainter than outbursts of AMXPs ($10^{36}-10^{37}\,\ergsec$) and brighter than rotation-powered MSPs and quiescent AMXPs ($10^{30}-10^{32}\,\ergsec$). {\psr1023}
has a {\it sub-luminous} disc for 7 years (and counting), with a little change of its
properties, if any. {\xss} behaved as such between 2003 and 2012, and
possibly since earlier times. 
Shorter episodes were also recorded in {\psr1023} and {\igr} (see Sec.\,\ref{sec:sources}).

\begin{figure}[t!]
\includegraphics[scale=0.67]{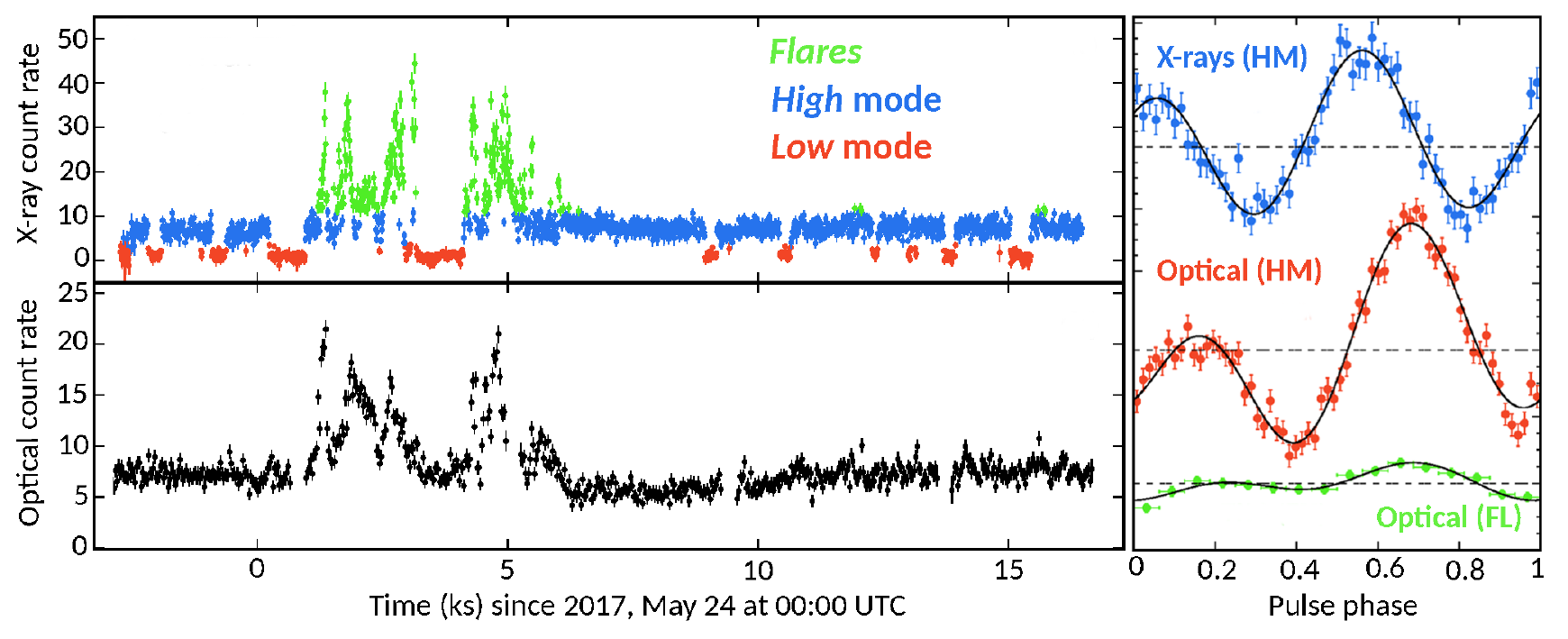}
\caption{X-ray (top-left panel) and optical (bottom-left panel) light
  curves of PSR J1023+0038 in the {\it sub-luminous} disc state
  observed simultaneously by {\it XMM-Newton}. {\it High} mode, {\it
    low} mode and flares are plotted in blue, red and green,
  respectively. The right panel shows the X-ray pulse profile during  the {\it high} mode 
  (blue points)  and the optical
  pulse profiles 
  in the {\it high} (red points) and flares
  (green points) detected by {\it SiFAP} at the {\it INAF Galileo
    Telescope}. Figure adapted from 
    \cite{pap19}. }\label{fig:j1023}       
\end{figure}


{\bf The intensity modes -- }The {\it high} (sometimes termed {\it
  active}) and the {\it low} ({\it passive}) intensity modes observed in the X-ray light curves together with sporadic flares, are perhaps the defining characteristics of tMSPs in the  {\it sub-luminous} state. The top-left panel of Fig.~\ref{fig:j1023} shows an X-ray light curve observed from {\psr1023}; the {\it high} and {\it low} modes are plotted with blue and red points, respectively, while flares are shown with green symbols.  Most of the information about these intensity modes has been obtained from observations of {\psr1023} \cite{pat14, cze14, ten14, bog15, cam16}, although they have also been observed from {\xss} \cite{sai09,dem10,dem13} and {\igr}
\cite{pap13,lin14} (see \cite{lin14b} for a comparative study). {\psr1023} lies for $\sim 80\%$ of the time in the {\it high} mode, emitting a roughly constant 0.5--10~keV
X-ray luminosity  of $\sim 3\times10^{33}$~erg~s$^{-1}$.
Unpredictably, sharp transitions to the {\it low} mode occurs on a timescale of $\sim 10$~s. The X-ray luminosity observed in the {\it low} mode is also roughly constant and about one order of magnitude fainter than in the {\it high} mode,  
but still a few times brighter than the
rotation-powered state (see Table\,1). The transition from the {\it low} to {\it high} mode are characterised by a similar timescale. The duration of these modes ranges from a few tens of seconds to a few hours, although there does not seem to be a characteristic length or recurrence time, nor a correlation between  waiting times, duration
or luminosity.  The luminosity of the {\it high}
and {\it low} modes have been stable within $\sim 10\%$ in
the many observations of {\psr1023} performed so far, without any modulation at the
orbital period.

\begin{figure}[t!]
\includegraphics[width=1.0\textwidth]{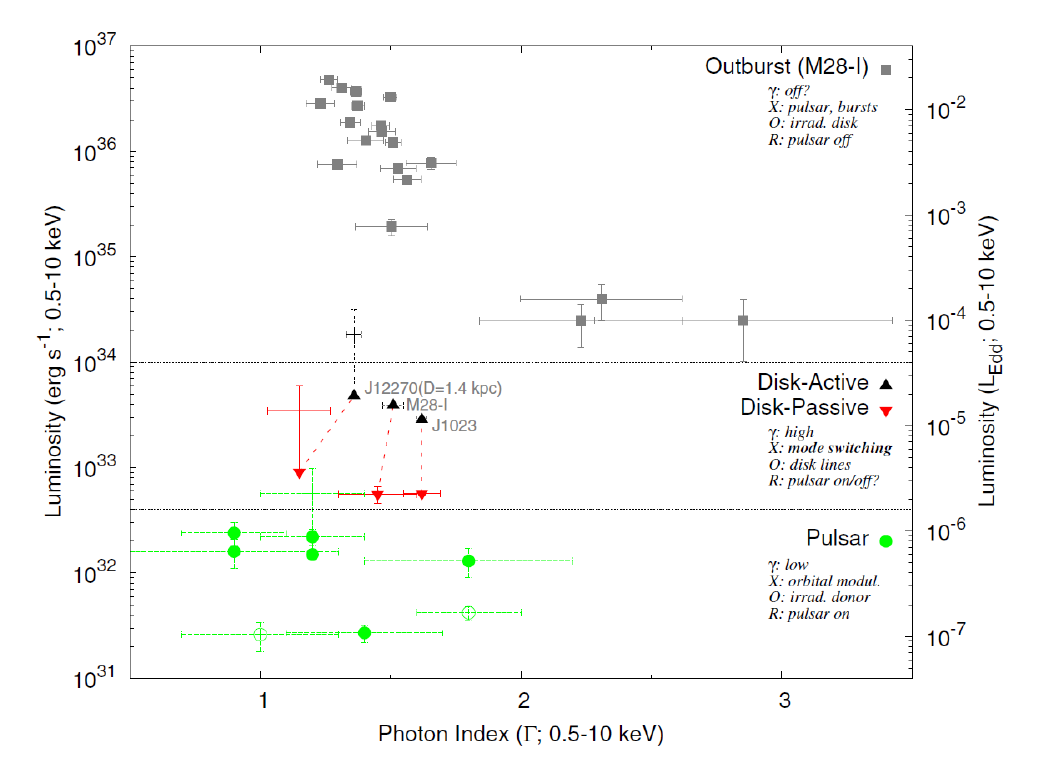}
\caption{0.5--10 keV X-ray luminosity and power-law spectral index of
  tMSPs 
  in outburst (M28I={\igr} )  (top panel), in the {\it high}/active
  and {\it low}/passive modes in the {\it sub-luminous} disc state
  (middle panel) and of tMSPs and redbacks in the radio pulsar state (bottom panel). Taken from
  \cite{lin14b}.}
\label{fig:lin14b}       
\end{figure}

The spectrum of the X-ray emission observed in the {\it high} mode was described by an absorbed power-law with an index of $\Gamma_X=1.4-1.6$,  significantly softer than in the radio pulsar state, $\Gamma\simeq1.1-1.2$
(\cite{lin14b,bog15,cze18}, see Tab\,1 and Fig.~\ref{fig:lin14b}). The power-law extends up to at least $\sim 80$~keV without evidence for a cut-off \cite{ten14}. A thermal component with a temperature of $\sim 130$~eV and 
contributing to a few per cent of the total flux was also detected at soft X-ray energies in {\psr1023} 
\cite{bog15}. Its properties are compatible with the emission coming from the inner rings of a disc truncated $\sim 20$~km from the pulsar \cite{cam16}. In the
{\it low} mode, the thermal component disappears and the power-law spectrum becomes slightly softer than in the {\it high} mode ($\Gamma\simeq 2.0$,   \cite{cam16,cze18}).

The X-ray modes observed from {\xss} had similar properties. However, the energy spectrum  observed in the {\it low} modes which occurred after a flare  was harder
than in the {\it high} mode \cite{dem10,dem13,lin14b}. The tail of the
flaring emission likely contaminated those spectra, indicating that
the flaring mechanism is independent of the {\it high}-to-{\it low}
mode transitions \cite{mir20}. An additional partial covering neutral absorber was required to model the spectra of these {\it low} modes,
suggesting a refilling of the matter reservoir close to the NS after a
flare \cite {dem10}. The {\it high} and {\it low} modes observed in
{\igr} had a similar luminosity ratio ($\sim 7$), but lasted
significantly longer (up to 20~hours) and showed slower transitions ($\sim
500-1000$~s), although shorter timescales could not be probed due to the
particular observing mode  \cite{lin14,pap13}.

Flat-bottomed dips were also observed  in high-cadence optical
observations of {\psr1023} \cite{sha15,sha18}. The ingress/egress times were slightly longer ($\sim 20$~s) than those of  X-ray {\it low} modes, whereas the duration was similar. A bi-modal distribution of the optical flux was
also found from a lower cadence {\it Kepler K-2} monitoring 
\cite{ken18}.  However, simultaneous optical/X-ray {\xmm} observations have not
revealed such dips in B-band data \cite{bog15,bag19}, possibly
because the optical {\it low} modes are energy dependent.
The lack of simultaneous detection of optical dips and X-ray {\it low} modes has prevented to establish the relationship between these phenomena, so far.
Flares and hints of flat-bottomed dips were also found in near-infrared
K$_s$-band photometric light curves \cite{sha18}. An enhancement of the  near-infrared emission, possibly a flare, was also observed right after an X-ray  {\it low}-to-{\it high} mode transition 
\cite{pap19}.  {\it Low} and {\it high} modes were much more evident
in the ultraviolet; they occurred simultaneously with the X-rays
displaying variations by $\sim25-30\%$ in both {\psr1023} \cite{jao19}
and {\xss} \cite{dem10,dem13}.

{\bf The gamma-ray brightening} of {\psr1023} when switching from the
rotation-powered to the {\it sub-luminous} disc state, and the gamma-ray dimming of {\xss} in the reverse transition, have been
certainly one of the most unexpected features of tMSPs\footnote{The M28 globular cluster to which IGR J18245-2452 belongs, hosts a population of gamma-ray emitting MSPs  \cite{wu13} that made a gamma-ray brightening difficult to detect.}. AMXPs in outburst (and in general
LMXBs) have not been detected in gamma-rays, so far\footnote{Note that AMXPs are generally farther ($d\simeq5$--$8\,$ kpc) than the two tMSPs in the Galactic field ($d\sim 1.5\,$ kpc).}. On the other hand, tMSPs in the {\it sub-luminous} state became a
few times brighter than in the rotation-powered state and slightly brighter than in the X-ray band  (see Tab.\,1). The $\sim$ tenfold gamma-ray
brightening observed from {\psr1023} in 2013 took
place in a month, or less \cite{sta14,tor17}. The transition of {\xss} in
the opposite direction was smoother and less pronounced 
($ L_{\gamma,disc} \sim 2.5\, L_{\gamma,rot}$) 
\cite{joh15,tor17}. The gamma-ray spectra of both tMSPs in the disc
state were well described by a power law with index
$\Gamma_\gamma\approx 2.0$ with marginal evidence of a cutoff between 4 and 10 GeV (\cite{tor17}, see Table\,1).  
A high-energy ($>5$~GeV) component was recently
claimed to emerge in the spectrum of {\psr1023} at orbital phases
corresponding to the pulsar descending node \cite{xin18}, but 
confirmation with a higher counting statistics is warranted. So far, only
upper limits have been set  to the emission in the TeV regime \cite{ali16}.

{\bf The radio emission -- } In the {\it sub-luminous} state of
{\psr1023}, radio (0.3-5\,GHz) pulsations 
have not been detected in either of the X-ray modes. Upper limits of
0.1-1~mJy were set, i.e. more than an order of magnitude lower than in the
radio pulsar state \cite{sta14,pat14,bog15}. As the  radio emission could be
absorbed by the intrabinary material, this does not necessarily imply the
complete quenching of the radio pulsar.

\begin{figure}[t!]
\includegraphics[width=1.0\textwidth]{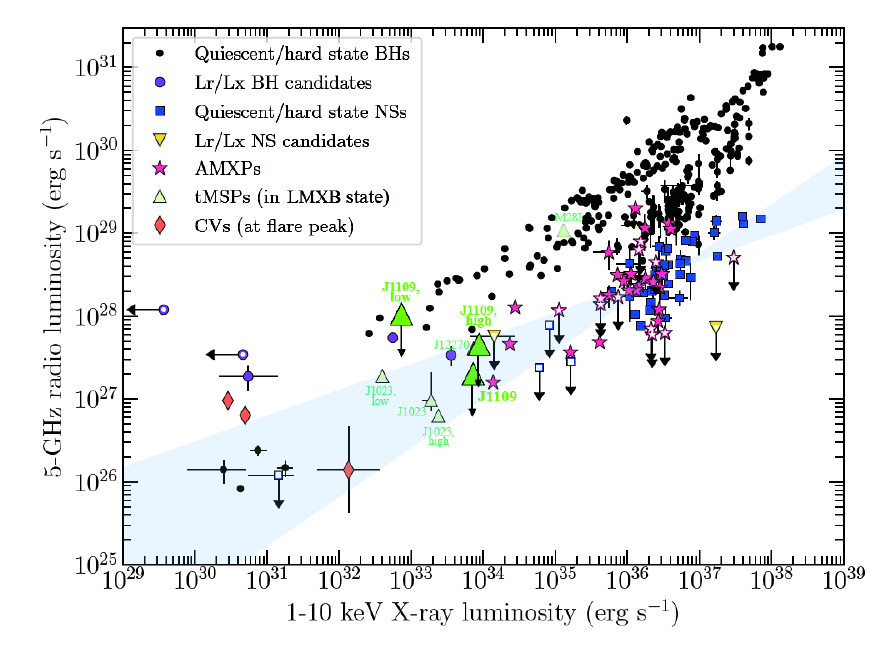}
\caption{Radio vs. X-ray luminosity plane for different classes of
  accreting compact objects. TMSPs in the {\it sub-luminous} disc
  state are plotted with green symbols. The cyan shaded area encloses
  the 3$\sigma$ confidence interval on the correlation holding for
  accreting NSs \cite{gal18}. The figure is taken from 
  \cite{cze19} and adapted from
  \cite{bah18}. }
\label{fig:cze19}       
\end{figure}

A radio continuum emission with a flat or a slightly inverted spectrum was instead seen from  both {\psr1023} and {\xss} \cite{hil11,del15}.  Similar  spectra are ubiquitous
among accreting X-ray binaries in the hard state and are interpreted
in terms of  partially self-absorbed synchrotron emission from
out-flowing material \cite{fen16}. Different correlations hold between
the X-ray and the radio luminosity of X-ray binaries hosting black
holes and NSs; the latter are generally fainter radio sources at a given X-ray
luminosity \cite{gal18}. In a $L_{\it radio}-L_{X}$ diagram (see
Fig.~\ref{fig:cze19}), tMSPs fall in the radio-bright end of the range
expected by propagating the correlation for bright accreting NSs to
a lower X-ray luminosity, especially when the the peak radio
luminosity is considered.

Simultaneous radio and X-ray observations of
{\psr1023} unveiled  an anti-correlated pattern of variability
\cite{bog18}. When the source switched from the {\it high} to the {\it
  low} X-ray mode, the radio flux suddenly increased and its spectrum
became steeper. The decay of the radio emission was instead shallower; it started earlier than the
{\it low}-{\it high} mode transition and ended $\sim 30-60$~s later.  This
phenomenology suggested that optically thin emission from expanding
plasmoids becomes dominant when the source drops into the {\it low}
mode. Sporadic radio flares up to a few mJy have also been 
observed in this source \cite{bon02}. In particular, a radio flare with evolving synchrotron features typical of accretion driven outflows was observed to occur a few minutes after a bright X-ray flare, although other radio flares were observed when the source was in the {\it high} X-ray mode
without any appreciable  variability \cite{bog18}.

{\bf The optical/UV emission --} The optical brightness of tMSPs
in the {\it sub-luminous} state is $\sim\,1-2$ magnitudes brighter than in the rotation-powered state, due to the contribution of the newly-formed accretion disc \cite{hal13,pal13,bas14}. On the other hand, the donor stars were still found to be heated at a similar level as in
the rotation-powered state (see Sec.~\ref{sec:radiopsr}).  The V and R
band emissions in {\psr1023} were found to be linearly polarised at
$\sim 1\%$ level possibly due to Thomson scattering within the disc
\cite{bag16,hak18}.

The optical spectra were dominated by a blue continuum and strong,
double-peaked emission lines of H and He produced in an optically
thick accretion disc \cite{wan09,dem14,cze14,dem14,bog15}. Modelling of
 the spectrum of {\psr1023} observed during the
2000-2001 accretion state with a simple disc model gave a temperature in the range 
$(2-34)\times10^3$~K, and inner and outer radii of $R_{in}\sim10^9$~cm
and $R_{out} \simeq 6\times10^9$~cm, respectively \cite{wan09}. The inner disc radius is larger
than the Alfven radius, indicating that the optical emission
originated in the outer disc regions. Similar results were obtained
modelling the UV spectrum  \cite{her16}. In {\xss},
Doppler tomography showed that the hotter regions  producing the He\,II
 emission lines were similarly far-out \cite{dem14}.

{\bf Flares} were    observed simultaneously in the X-ray, UV, optical,
and near-infrared bands both from {\psr1023} \cite{bog15,hak18} and
{\xss} \cite{dem10,sai11,dem13,dem14}, suggesting a common underlying
process. X-ray flares emitted most of the energy ($L_{X}\simeq 6
\times L_{opt}$). They  lasted from less than a minute to a
few hours \cite{dem13, bog15,jao16}, even though extended episodes
lasting up to ten hours have been observed in {\psr1023}
\cite{ten14}. The brightest observed flares attained an X-ray luminosity of
$\approx 2-7\times 10^{34}\,\ergsec$ in the 0.3--79~keV band,
(\cite{ten14,bog15}, see Table\,1), slightly exceeding the pulsar
spin-down power.

The most prominent optical flares of {\psr1023} had amplitudes of
$\sim 0.5-1$ mag and lasted up to 14~hours; they traced the brightest
flares seen in X-rays with both positive and negative lags of up to
$\sim 250$~s \cite{bog15}. An 80-days long {\it Kepler K2} coverage observed optical flares for 15-22$\%$ of
the time (depending on the flare identification algorithm,
\cite{pap18,ken18}), much more frequently than X-ray flares  observed at other epochs ($\simeq 2$ per cent, \cite{jao16}). This
showed that the flaring activity is highly unpredictable and cannot
be easily parameterised.  Multi-band simultaneous observations of a
bright event suggested that the flare emission became hotter and more optically
thin, like in an accretion disc corona and/or  hot
fireball ejecta \cite{sha15}. Similar indications were also found in
{\xss} \cite{dem10}. Remarkably, the optical emission lines observed from {\xss} \cite{dem14} and {\psr1023} \cite{hak18}  showed a
tendency to disappear during  intervals characterised by enhanced
flaring emission,
compatible with the onset of an outflow. The emergence of an
additional polarised component during the flares, possibly due to Thomson
scattering from ejected matter, also supported this hypothesis
\cite{hak18}.  Flaring variability observed in the near-infrared 
lagged the optical variability by $\sim 10$~s, and was tentatively
attributed to the reprocessing of the optical emission produced close to the
light cylinder by a stream of matter ejected by the system further out
\cite{bag19}.

{\bf Pulses --} X-ray pulsations at the NS spin period were detected
only during the {\it high} mode in both {\psr1023} (\cite{arc15}, see
the right panel of Fig.~\ref{fig:j1023}) and {\xss} \cite{pap15}. They
had an rms amplitude of $\sim 6-7$ per cent and were modelled with two
sinusoidal harmonic components. Pulsations instead disappeared during the
{\it low} modes and the flares, with upper limits of $\sim 1-2\%$.  A quasi-coherent timing solution measured over the interval Nov 2013 - Dec 2015 remarkably found that {\psr1023} was spinning down in the
disc state at a rate $(32\pm2)\%$ larger than in the radio pulsar
state \cite{jao16}\footnote{The fractional change reported here is slightly larger than the value quoted in \cite{jao16}, $(26.8\pm0.4)\%$, because it takes into account the Shklovskii effect and acceleration in the Galactic potential \cite{del12}.}.

Quite surprisingly, also the optical emission observed from {\psr1023}
turned out to be pulsed with an rms
amplitude of $\simeq 1\%$ \cite{amb17,pap19,zam19,kar19}. This made {\psr1023} 
the first optical MSP to date. Optical pulsations were
detected in the {\it high} mode and disappeared in the {\it low} mode with
an upper limit of $A<0.034\%$ (\cite{pap19}; see the right panel of Fig.~\ref{fig:j1023}). Optical and X-ray pulsations had a similar shape with the optical
lagging the X-rays by $\sim (200\pm70)\,\mu$s, an estimate affected by large systematics. The spectral energy
distribution of the pulsed emission from the optical to the X-ray band
was found to nicely match a power-law relation $F_{\nu}
\propto{\nu}^ {-0.7}$ (see the left panel of
Fig.~\ref{fig:j1023sed}). All these properties  indicate that the optical and the X-ray pulsations share a common underlying mechanism \cite{pap19}.
  Interestingly, the optical pulses were also
detected during flares with an amplitude $\sim$ six times lower than
that in the {\it high} mode \cite{pap19}. The spin-down rate determined from optical pulsations measured between Jan 2018 and Jan 2020 was $\sim 20\%$ lower than that measured from X-ray pulses at earlier epochs, 
and closer to the value observed in the radio pulsar state \cite{bur20}. The difference in the spin-down rate measured from the optical and the X-ray pulsations at different epochs has to be investigated with further long-term coverage in both domains.

\subsection{Candidate transitional millisecond pulsars}
\label{sec:cand}

Stimulated by the discovery of the three tMSPs, searches for new
candidates started just after. Searching for a counterpart to a yet unidentified gamma-ray source, with peculiar time variability (such as the {\it high-low} modes and flares) and spectral properties in the optical (e.g., double peaked emission lines and a blue continuum) and X-ray bands  ($L_X(0.3-79\,\mbox{keV})\simeq (0.5-1) \times L_{\gamma}(0.1-300,\mbox{GeV})$ and a power law-shaped spectrum with a photon index $\Gamma\sim 1.7$), turned out to be the most efficient way to identify candidate tMSPs in the {\it sub-luminous} disc state. 

{\bf RXS\,J154439.4-112820} was found within the error
circle of {\3fgl}. The spectrum of the optical counterpart showed
prominent H and He emission lines, consistent with the presence of an
accretion disc \cite{mas13}. The X-ray light curve
displayed  {\it high} and {\it low} modes differing
by a factor of $\approx 10$ in flux, with transitions occurring on a timescale of
$\approx 10$~s \cite{bog15b}. The power-law shaped X-ray spectrum, featuring a
photon index $\Gamma\simeq$1.7 and marginal  spectral
variability, was also similar to those shown by tMSPs in the {\it
  sub-luminous} disc state \cite{bog16}.  The long-term optical light
curve revealed variability by 0.5~mag and in some occasions
enhancements by $\approx$1.0-1.5\,mag \cite{bog16}, reminiscent of the
optical flares observed in tMSPs. Optical time-resolved spectroscopy
measured a period of 5.8\,hr and an inclination of 5-8$^{\circ}$,
implying that the system is seen almost face-on \cite{bri17}. Adopting
a distance prior to the {\em Gaia} DR2 parallax \cite{gai18}, based on
the Galactic model by \cite{bai18}, gave a distance of
1.87$_{-0.72}^{+1.9}$\,kpc. The corresponding X-ray and gamma-ray
luminosity are $\rm \simeq 3.6\times10^{33}\,erg\,s^{-1}$ and $\rm
5.9\times10^{33}\,erg\,s^{-1}$, respectively, very similar to those
of tMSPs in the {\it sub-luminous} disc state.

 {\bf CXOU\,J110926.4-650224} (also known as  IGR J11098--6457 \cite{tom09}) is a hard X-ray source located within the error circle of FL8Y\,J1109.8--6500. The typical  modes shown by tMSPs in the {\it
   sub-luminous} disc state were easily recognised in its X-ray light
 curve, and also the energy spectrum was similar to those of the tMSPs  \cite{cze19}.  Its variable optical counterpart showed disc emission lines, whereas shallow radio continuum observations could only set  an
 upper limit, still compatible with typical tMSP characteristics.  At
 a distance of $\sim4$ kpc, derived from {\em Gaia} DR2, the X-ray
 luminosity is $\rm \simeq 2.2 \times 10^{34}\,erg\,s^{-1}$, of the same order of  the gamma-ray luminosity $\rm \simeq 1.5\times10^{34}\,erg\,s^{-1}$
 \cite{cze19}.

 Recently, a variable optical and X-ray source has also been found within the error circle of the gamma-ray source {\bf 4FGL~J0407.7--5702} \cite{mil20}. The optical spectrum showed double peaked H and He emission lines and a blue continuum indicating the presence of an accretion disc.  The X-ray spectrum is also found to be compatible to those observed in tMSPs.   The ratio of the X-ray to gamma-ray luminosity was similar to that shown by other sources in the {\it sub-luminous} disc state, making it a strong candidate tMSP. The comparison of the flux observed in the optical, X-ray and gamma-ray bands with those of other tMSPs in the disc state, and the lack of a significant Gaia parallax, have suggested a distance larger than 5~kpc. This would make it the farthest tMSP known. Although  the distribution of the count rates observed in the X-ray band featured two peaks, the phenomenology of the light curve  makes  an interpretation in terms of the  usual {\it high}--{\it low} modes somewhat  difficult although the time spent in one or other mode can be different from source to source.

 The gamma-ray source {\bf 3FGL J0427.9-6704} was recently associated with an accreting
 eclipsing X-ray binary with an $8.8$ hr period and a hard X-ray spectrum extending up to $\sim 50$~keV \cite{str16}.   The eclipses were 
 observed  both in  the X-ray and the gamma-ray light curves, demonstrating the
 association of the counterpart, and indicating that the high energy emission arose very
 close to the compact object \cite{str16, ken20}. The bright and variable emission of the radio counterpart was instead not eclipsed, indicating an origin
 further out. 
  The 
 X-ray luminosity
 of this source ($\rm 5\times 10^{33}\,erg\,s^ {-1}$ at 2.3\,kpc) was of the same order of that that
 observed at gamma-ray energies, similar to tMSPs in the {\it sub-luminous} state. However, the simultaneous X-ray and optical/ultraviolet light
 curves did not reveal the typical intensity modes of tMSPs, as the
 source was found to be flaring for most of the time
 \cite{li20}. Modelling of the optical orbital modulation and of the radial velocity of lines originating from the surface of the  $\rm M_2 \sim
 0.6\,M_{\odot}$ companion star indicated a relatively massive $\sim 1.8-1.9$~M$_{\odot}$ NS \cite{str16}. However, a recent study based on much
 higher quality photometric data found a lower, although
 not highly constrained, mass for the primary
 ($1.43^{+0.33}_{-0.19}\rm M_{\odot}$,  \cite{ken20}). Further X-ray observations
 aimed at detecting the typical modes of tMSPs will constrain the true
 nature of this candidate.

These characteristics listed above  
make these sources  very strong candidate tMSPs in the {\it sub-luminous} disc state,
 although the lack of precise orbital ephemeris  has hampered the detection of X-ray
 pulsations, so far.

The recent release of the $\rm 4^{th}$ {\em Fermi}-LAT catalogue \cite{abd20} and its newest 10-yr DR2 version\cite{bal20} \footnote{https://fermi.gsfc.nasa.gov/ssc/data/access/lat/10yr\_catalog/} enhanced the search for MSP binaries associated with an X-ray and optical
counterpart. However,  discriminating
between redbacks and tMSPs in the disc state has been sometimes not 
immediate. For instance,  a few redbacks such
as {\j0838} \cite{rea17,hal17}, PSR\,J1048+2339 \cite{str19} and PSR\,J1628-3205
\cite{cho18,str19}, the black widow PSR\,J1311-3430 \cite{rom15} and
the recently discovered long-period ($>1$\,d) MSP binaries
2FGL\,J0846.0+2820 \cite{swi17} and 3FGL\,J1417.5-4402
\cite{str16,swi18} occasionally displayed emission lines in
their optical spectra. While they may hint to a disc origin, the
emission lines could be also ascribed to a magnetically driven wind of
the companion. The recently identified source 4FGL\,J0935.3+0901 has
an optical counterpart with double-peaked emission lines, features gamma-ray
properties similar to tMSPs, and showed an enhancement by a factor
about 8 between Dec. 2010 and Jul. 2013 with significant spectral
change \cite{wan20}. However, its X-ray-to-gamma-ray flux ratio
($\sim 40$) was more typical of spiders in the rotational-powered
state rather than of tMSPs in the {\it sub-luminous } disc
state. Simultaneous photometric and spectroscopic observations will be
crucial to understand the connection of variable heating and the
appearance of emission lines.

\smallskip
Recently, very faint persistent or quasi-persistent
X-ray binaries with  a luminosity $\rm \sim 10^{33}-10^{35}
\,erg\,s^{-1}$,  have also been  proposed to
harbour tMSPs in the {\it sub-luminous} disc state \cite{hei15}. These sources would switch on as radio pulsars as soon as the X-ray luminosity drops below $\rm
\sim 10^{32}\,erg\,s^{-1}$. However, given the relatively low X-ray luminosity involved, detecting a
state transition  is  only possible for close-by sources or  deeply observed
fields, such as the Galactic Centre and globular clusters.

{\bf Terzan 5 CX10} (CXOGlb J174804.5-244641) is a variable hard X-ray
source in the dense globular cluster Terzan 5. A comprehensive study
of several {\em Chandra} observations spanning 13 years found it twice in
a bright state with $\rm L_X \sim 2\times 10^{33}$\,erg\,s$^{-1}$
(in 2003 and 2016), and twice in a much fainter state $\rm L_X \sim
10^{32}$\,erg\,s$^{-1}$ (between 2009 and 2014) with a harder spectrum
than the bright state \cite{bah18}. This behaviour was
 reminiscent of the changes of state of tMSPs.  A
faint optical counterpart with colours  compatible
with the cluster main sequence was also identified. Follow-up radio continuum observations
also revealed a faint ($\sim 20\,\mu$Jy at 3\,GHz) radio source, which
placed CX10 close to the
position of {\psr1023} in the X-ray/radio luminosity diagram. These properties make it a strong candidate
tMSP to be searched in beamed-formed radio observations to reveal the
yet to be discovered pulsar.

{\bf XMM J174457-2850.3} is a faint X-ray transient in the Galactic
centre region. The detection of a 2\,hr-long type-I X-ray burst proved
that it hosts an accreting NS \cite{deg14}.   This source exhibited  a few-weeks
long outbursts up to $\sim 10^{36}\,\ergsec$, but for most of the time, it lay in quiescence with a luminosity of $\sim 5 \times 10^{32}\,\ergsec$. Also, it was occasionally found
to linger for several months at an intermediate level of
$10^{33}-10^{34}\,\ergsec$ \cite{deg15}.  The X-ray spectrum was
described by a $\Gamma \sim 1.4$ power-law, much harder than that generally
observed from LMXBs at the same luminosity level.  The properties of these three luminosity states  resembled those  observed in the tMSP IGR\,J18245-2452 \cite{pap13}. However, no
meaningful search for fast pulsations could be performed either in
X-rays due to the low statistic of available data \cite{deg15}, or in
the radio band due to the large (6.5\,kpc) distance of the source.

Recently, a catalogue of more than 1100 X-ray sources in 38 globular clusters has been compiled to complement the  MAVERIC (Milky-way ATCA VLA Exploration of Radio-sources In Clusters) radio survey \cite{bah20}. Among these, a source in {\bf NGC 6539} was identified as a candidate tMSP based on its X-ray properties, and the presence of a bright radio counterpart with a flat/slightly inverted spectrum.    However, this source  occupies the same region in the  L$_X$-L$_{\rm radio}$ diagram of black holes and AMXPs (see Fig. 7) and further observations are required to address its true nature.

\section{Models and open questions}
\label{sec:models}


\subsection{The rotation-powered state}
\label{sec:radiopulsarmodels}

The relativistic wind of MSPs in close binaries is terminated by the interaction with the stream of matter issuing from the companion star or  the companion star itself. Hence, they offer the opportunity to study the properties of the termination shock so created at much smaller distances than in pulsar wind nebulae. Already in late-80ties, it was also predicted that high-energy photons generated by the particles accelerated at the termination shock would have been able to evaporate the late-type companion star  \cite{klu88,rud89,tav91}.  Models were first applied to the case of the first black window pulsar discovered, PSR\,B1957+20 \cite{aro93}. Given the
relatively small size of the binary ($d\simeq 10^{11}$~cm), the
magnetic field down-stream the shock is  $B\simgt 3
\sqrt{L_{sd}/c d^2} \approx 30$ ~G (for a magnetically dominated and
isotropic wind emitted by a pulsar with spin-down power $L_{sd}\simeq
10^{34}\,\ergsec$). Synchrotron emission is thus the main cooling mechanism of the relativistic particles accelerated in the shock, yielding an X-ray output which exceeds the
magnetospheric pulsar emission. A recent X-ray study
of a large sample of MSPs indeed found that redbacks are
brighter than black widows and isolated MSPs \cite{lee18}. This indicates that a larger fraction of the pulsar wind of redbacks is intercepted at the shock surface compared to black widows. 

\noindent The luminosity of the shock synchrotron emission depends on the strength of the magnetic field beyond the shock. Assuming that the  efficiency of electron acceleration at the shock is similar to that of the Crab pulsar, the relatively bright X-ray luminosity observed from tMSPs required a pulsar wind dominated by the electromagnetic Poynting flux and focused along the equatorial plane of the pulsar (expected to be close to the orbital plane for a spun-up MSP) \cite{bog11,bog14,dem20}. The
electron population has a power-law energy spectrum with an index $p$
which is related to the power-law index $\Gamma$ of the X-ray spectrum as $p
\sim 2\Gamma$ - 1.  The X-ray spectrum of both tMSPs \cite{ten14,dem20} and other redbacks
observed with {\em NuSTAR} \cite{kon17,aln18,kan19}
extended up to at least 70\,keV, and was consistent with a power-law
with index $\Gamma \sim 1.1-1.2$, implying $p\sim 1.3$. Such a value
favours a shock-driven magnetic reconnection in a striped pulsar wind
(see \cite{sir15} and reference therein) rather than diffusive shock
acceleration. Even though the shock emission has to extend well above
the 3--79~keV hard X-ray band covered by {\em NuSTAR} to be
efficient enough in irradiating the secondary star \cite{dem20}, it must be
limited below a few MeV not to exceed the pulsar spin-down power.
The energy dependence of the orbital modulation marginally seen
in two tMSPs (see Sec.~\ref{sec:radiopsr}) and PSR\,J2129-0429
\cite{aln18} could also hint at a spatial variation of the $p$-index of the
synchrotron emitting electrons.

\noindent The prediction that the X-ray emission was
modulated at the binary orbital period due to the
obscuration by the companion star and to Doppler boosting  \cite{aro93} was indeed  confirmed
in several MSP binaries \cite{bog05,rob18}.  The phasing of the
X-ray orbital modulation observed in tMSPs is similar to that observed in
other redbacks; the X-ray flux attains a maximum when the pulsar is at the inferior
conjunction of the orbit, in phase with the optical orbital variability
\cite{rob18}. On the contrary, the X-ray modulation of black widows
displays a minimum when the pulsar is at the inferior conjunction of the orbit. 
To explain this, a different orientation of the shock in redbacks and black widows has been
proposed by two groups \cite{rom16,wad17,wad18}, who have developed
semi-empirical models to explain the radio, X-ray and optical
behaviour.  They argued that the shock which surrounds redbacks is
oriented towards the pulsar  due to the large
companion wind momentum $\beta_w = \dot M_2\, v_w\,c /\dot E$ (where
$v_w$ is the relative wind velocity and $\dot M_2$ is the mass loss
rate), while it surrounds the companion in  black widows. The wind
momentum ratio also sets the shock opening angle, which together with
the binary inclination in turn determines the shape of the X-ray orbital
light curve, single or double-peaked; the width of the peaks depends
instead on the boost parameter \cite{dub15,wad17}.  Most redbacks
showed a rather stable X-ray double-peaked modulation, while the tMSP 
PSR\,J1227-4853 displayed variations from single to double and again
single-peaked shape over several months, indicating changes in the
shock parameters \cite{dem15,dem20}. It was also noticed that systems
prone to make or just after a transition, may indeed display variability
in the shape of X-ray orbital modulation \cite{wad17}. 

\noindent  The stability
of the shock over years is still an unresolved issue since
a quasi-radial in-fall terminated outside the pulsar light cylinder is
unstable on dynamical timescales (\cite{bur01}; see
Sec.~\ref{sec:changes}). It was suggested that either a highly
magnetised ($B\sim$ several kG) donor star with and low mass-loss rate
($\lesssim 10^{15}$~g~s$^{-1}$, \cite{arc13}), or a secondary star with a large mass
loss rate but with an ADAF-like or heating-dominated flow, could bend
the shock towards the pulsar helping make it stable \cite{wad18}.
This flow should be unmodulated and detectable at soft X-rays down to
UV wavelengths and could explain the observed UV excess in
PSR\,J1227-4538.

\noindent The companion star heating pattern inferred from high-quality
optical photometric  light curves of both redbacks and black widows did not match
what expected from direct irradiation by the pulsar only, requiring also the 
illumination by the intrabinary shock \cite{rom16}.  An additional
source of heating could arise if a fraction of the wind particles
threads the companion field lines and is ducted to its surface; this would require a very active magnetic star
displaying star-spots or flares
\cite{san17,wad18}.  This possibility was claimed to
explain the optical light curve of the strongly irradiated companion
of the redback PSR\,J2215+5135 \cite{san17}. Asymmetries in the optical orbital
modulation were also observed in the tMSP PSR\,J1227-4853, although no
indication of a magnetically active star was found \cite{dem15}.

\subsection{The accretion-disc state}
           \label{sec:accretionmodels}

Transitional systems have shown a marked preference for the {\it
  sub-luminous} disc state than the bright X-ray outbursts typically
seen from AMXPs. This made them more elusive to discover, and hard to
reconcile with the typical classification scheme of X-ray
transients. 
The
main features of the {\it sub-luminous} state to explain are:
\begin{itemize}
\item its duration (more than $\sim$10 years) and faintness (the
  accretion rate estimated from the X-ray luminosity is
  $5\times10^{-5}$ times the Eddington rate);
\item the {\it high} and {\it low} intensity modes with fast ($\sim 10$~s)
  transitions seen in X-rays and UV, as well as in the optical and
  near-infrared, although with a still uncertain relationship with the
  other bands;
\item the X-ray and optical pulsations detected in the {\it high}
  mode;
 \item  the spin down of the NS, at a rate somewhat higher than in the rotation powered state;
\item a radio brightening occurring simultaneously to the X-ray {\it low}
  modes and  attaining a  luminosity comparable to black hole binaries in the hard
  state at the same X-ray luminosity;
\item an increased gamma-ray emission ($L_{\gamma}\simgt L_X$) compared to the rotation-powered state;
\item flares seen in
  the X-rays, UV, optical and near-infrared bands, with duration ranging from  several minutes to hours.
\end{itemize}
Determining whether the multi-wavelength
emission observed in the {\it sub-luminous} disc state is accretion or
rotation-powered is the major challenge.  This is not surprising
since the accretion luminosity estimated from the X-ray flux is
comparable to the pulsar spin-down power ($\approx \mbox{few}
\times 10^{34}\,\ergsec$), and both processes should be important.  
 Most of the models proposed so far relied on the standard assumption that the source emission could be either accretion or rotation-powered. In the accretion-powered case the intrusion of high-density accreting plasma into the magnetosphere would easily suppress the acceleration of particles in the magnetosphere and the resulting emission \cite{shv70,shv71,lip87}. On the other hand, the switch on of a rotation-powered radio pulsar would develop a radiation pressure which is able to eject the material lost by the  companion  \cite{bur01}. However, the  complications in applying one or the other assumption to the {\it sub-luminous} state of tMSPs have forced to consider models in which both rotation and accretion-powered mechanisms conspire to yield the puzzling emission properties listed above. 

{\bf Enshrouded radio pulsar models} - The unexpectedly bright
gamma-ray emission of tMSPs first led Takata et
al. \cite{tak14,li14} and Coti Zelati et al. \cite{cze14} to 
argue that a radio pulsar was hiding behind the enshrouding 
intrabinary matter \cite{tav91}. They assumed that the pulsar wind
truncated the disc far from the pulsar ($d \approx
10^9-10^{10}$~cm). The electrons accelerated in the shock would up-scatter the disc UV photons to yield the observed
gamma-rays. These electrons would also interact with the field
permeating the shock to emit synchrotron X-ray photons.

\noindent These models were proposed before the optical and  X-ray pulsations had been discovered. Although the magnetosphere of a rotation-powered pulsar could produce these pulsations, 
the efficiency in converting the spin-down power into optical and
X-ray pulsed emission ($\mbox{a few}\times 10^{-4}$ and
$6\times10^{-3}$, respectively for {\psr1023}) should be higher than in young
rotation-powered optical (see Fig.~3 in \cite{amb17}) and X-ray
pulsars (see \cite{lee18}).  Also, the X-ray
efficiency should have increased by $\sim 25$ times after the
formation of the disc. This would be hard to understand since the magnetospheric processes of the pulsar should not be affected by a disc truncated much further out.  Besides,
the synchro-curvature models which provided a successful modelling of the
X-ray/gamma-ray emission of other MSPs \cite{tor18,tor19}, failed to do so
for tMSPs in the disc state.

{\bf Accretion/propeller models - }The detection of X-ray pulsations
with similar properties of the pulses of AMXPs suggested that accretion onto the NS
magnetic poles was  taking place also in tMSPs \cite{arc15,pap15}.  However,
this would make tMSPs the faintest accreting X-ray pulsars known. This
is a critical issue since the mass accretion rate deduced from the
observed X-ray luminosity ($\dot{M}\simeq
5\times10^{13}$~g~s$^{-1}=5\times10^{-5}\,\dot{M}_{Edd}$) would place
the accretion radius well beyond the co-rotation radius
(e.g. $R_{acc}\sim 75$~km and $R_{co}\simeq25$~km in {\psr1023}, see
eq.s~\ref{eq:racc} and \ref{eq:rco}). A centrifugal barrier would be
expected to inhibit completely the accretion inflow \cite{ill75}.
Magneto-hydrodynamic simulations \cite{rom18} have shown that if the
magnetosphere rotates only slightly faster than the disc matter
($R_{acc}\simgt R_{co}$), the propeller is {\it weak}; part of the
in-flowing mass manages to accrete and produce X-ray pulsations, 
the rest is bounced back to the disc in a non-collimated wind. Therefore, various
attempts have been made to keep the accretion radius close to
co-rotation at such a low $\dot{M}$, such as considering a high NS magnetic dipole inclination \cite{boz18}.

\noindent Papitto et al.  \cite{pap14,pap15b}  argued instead  that  the mass
accretion rate in the disc was higher than that deduced from the X-ray
luminosity, so maintaining the accretion radius obtained with
Eq.~\ref{eq:racc} close to the co-rotation surface.  The propeller effect would eject most (>90\%) of
the disc mass with a low
emission efficiency, and only a tiny fraction would make its way to
the NS surface. Electrons would be accelerated at the magnetised ($B\sim
10^{5}-10^{6}$~G)  
turbulent disc/magnetosphere boundary and emit X-ray synchrotron photons.
The  Compton
up-scattering of these photons up to a few GeVs in a few
km-wide region would account for the gamma-ray emission.

\noindent Following D'Angelo et al. \cite{dan10,dan11,dan12}, it was alternatively
proposed that the disc of tMSPs could be trapped in a low
$\dot{M}$ state, so avoiding the onset of a propeller
\cite{jao16}. The in-flowing matter would pile up at the co-rotation boundary rather
than being ejected from the system, and the disc truncation radius
would be locked close to the co-rotation boundary without any strong
dependence on $\dot{M}$ (see also \cite{ert18} who obtained a similar
result in the propeller framework).

\noindent The transitions between the  {\it high} and {\it low} intensity modes could be due to a switching 
between an accretion/propeller and a rotation-powered state, respectively
\cite{lin14,cam16}. In the {\it low} mode, the pulsar wind would be terminated in a shock beyond the light cylinder, which would hide the radio pulses
\cite{cze14} and produce the power-law shaped 
X-ray spectrum. In the {\it high} mode, the disc would
 get close to the co-rotation radius, with most of the
emission produced at the boundary between the disc and the
propelling magnetosphere \cite{cam16}. The penetration of the disc
within the light cylinder would force some magnetic field
lines to open \cite{par16}, explaining the  enhanced
spin-down observed in the disc state  compared  to the radio
pulsar state  \cite{jao16,par17a,bur20}.

Bhattacharyya \cite{bha20} has recently included such an additional spin-down component in the torque budget. The fraction of magnetic field lines opened by the disc intrusion inside the light cylinder was estimated from the observed increase of the $\gamma$-ray emission. The underlying non-standard assumption was that the  magnetospheric processes invoked to explain the high energy emission in the  rotation-powered state kept working in the disc state, even if the in-falling plasma was accreting onto the NS surface so driving the X-ray pulsations. Since the accretion torques are negligible compared to the pulsar spin-down torques, the overall budget could be ensured only by the  inclusion  of an additional spin-down torque. Such a component could be granted by the continuous gravitational radiation related to a permanent  ellipticity of the NS  yielding a quadrupole moment of $Q\simeq 1-2 \times 10^{36}$ g cm$^2$. 

The increased spin-down due to the emission of gravitational waves had been first considered by Haskell \& Patruno \cite{has17}. They assumed that asymmetries in pycno-nuclear reactions or an unstable r-mode developing in the accretion state only could yield a  NS mass quadrupole moment of $>4.4\times10^{35}$ g cm$^2$, strong enough to account for the observed increase of the spin-down. Deriving  fully coherent X-ray and optical pulse timing solutions in the disc state, and a  radio pulse timing solution at the onset of the next radio MSP phase will test this intriguing possibility, which predicts a gravitational wave amplitude in the range of the next generation interferometers such as the Einstein telescope.

  Alternatively, Ertan \cite{ert18}  argued  that the sub-luminous disc state and the radio pulsar state represented weak and strong propeller states (although there is no strong evidence of the presence of a disc extending close to the NS when radio pulses are observed); a decrease of the accretion radius in the weak propeller state would account for the increased propeller torque.


\begin{figure}[t!]
\includegraphics[scale=0.55]{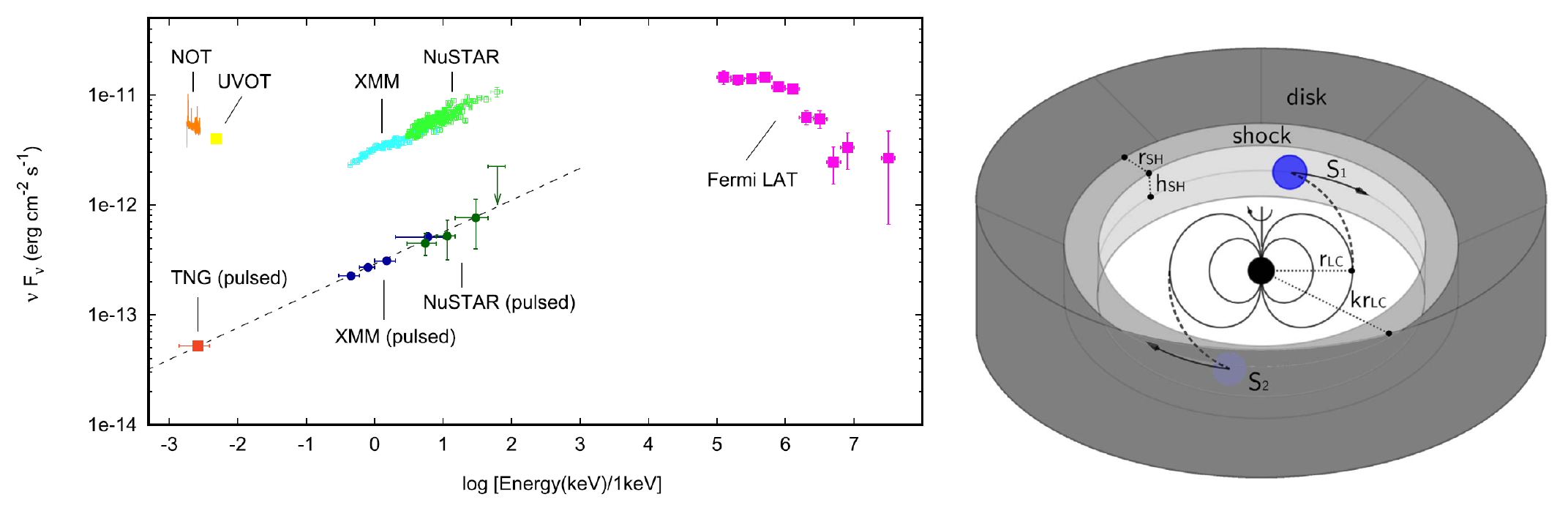}
\caption{Spectral energy distribution of {\psr1023} in the {\it
    sub-luminous} disc state (left panel). Sketch of the mini pulsar-wind nebula   configuration, taken from  \cite{pap19} (right panel).}\label{fig:j1023sed}

\end{figure}

{\bf A mini pulsar-wind nebula -- } The detection of relatively bright
optical pulsations from a tMSP in the disc state was hard to fit in
the accretion framework \cite{amb17,pap19}. The cyclotron self-absorbed
optical emission produced in the accretion columns is expected to be much fainter
than the values observed. A beaming of the emission by a factor of $\sim 50$ would
be required to match the observations, but it seems unlikely given
the sinusoidal shape of the pulses. The simultaneous appearance and
disappearance of optical and X-ray pulsations, the similar shape and the possibility of describing the
optical/X-ray pulsed spectral energy distribution with a single
power-law, strongly suggested that they are produced by a common underlying process.  As a consequence also the accretion
interpretation of the X-ray pulses had to be questioned. Papitto et
al. \cite{pap19} and Veledina et al. \cite{vel19} proposed that the
disc was truncated just beyond the light cylinder also in the {\it
  high} mode and that both optical and X-ray pulsations originated at
the pulsar wind termination shock. In the so-called striped wind
models (see, e.g., \cite{bog99}), two current sheets carry the
electromagnetic power of the pulsar wind outside the light
cylinder. These would produce two rotating spots in the inner face of
the wind/disc boundary in which particles are accelerated and quickly
radiate  optical and X-ray synchrotron photons, by interacting with the
relatively strong field ($\approx \mbox{few}\times10^5$~G) that
permeates the shock. An observer would see these spots from a different angle at each rotational phase, so explaining the detection of coherent optical and X-ray pulsations (see the right panel of Fig.~\ref{fig:j1023sed}). Indeed, the narrow emission
and absorption lines observed at a few keV supported the presence of a
hot and dense turbulent medium close to the light cylinder
\cite{cze18}.

\noindent In the {\it low} intensity mode, the termination shock would be pushed outward and the pulsations would be smeared because the synchrotron emission time scale and  the light travel time between different
regions of the shock become longer
\cite{pap19}. In the {\it high} mode, the shock would instead approach
the light cylinder, justifying the need of an additional absorbing
component covering 30\% of the emitting source to explain the change
of the X-ray emission spectrum compared to the {\it low} mode
\cite{cam19,mir20}.  Alternatively, the {\it low} mode could be
ascribed to the penetration of the disc plasma inside the light
cylinder which would curb the termination shock emission \cite{vel19}.

\noindent Axisymmetric general-relativistic MHD simulations demonstrated that
the pulsar electromagnetic wind can keep the plasma inflow beyond the
light cylinder, creating a termination shock and inhibiting mass
accretion \cite{par17}. Placing the termination shock close to the
light cylinder would also help solve the stability issue of the
equilibrium between the pulsar wind and the matter inflow beyond the
light cylinder \cite{eks05}. In this framework, episodes of
magnetic reconnection at the termination shock, in the disc or from
the donor star could also explain the observed flares \cite{cam19}.

{\bf Outflows -- } The bright continuous radio emission \cite{fer14,del15,bog18}
and the obscuration of the disc emission lines at certain orbital phases
\cite{dem14} suggested that tMSPs  launch outflows of plasma.
The radio emission observed in the {\it high} mode is compatible with self-absorbed synchrotron emission from a compact jet, whose spectral
break would be beyond the near-infrared band given the low accretion
luminosity \cite{bag19}. Less collimated outflows could  also be
launched by the propelling magnetosphere \cite{pap14,pap15b} or by
the pulsar wind \cite{pap19}.

\noindent On the other hand, a compact jet could hardly  explain the radio brightening observed in the {\it low} X-ray intensity mode, since a  correlation between the radio and X-ray luminosity
would be expected. 
The sudden radio brightening at the onset of an X-ray {\it low} mode
indicated that the emission had to come from the vicinity of the
NS. The launching of optically thin plasmoids by episodes of magnetic
reconnection due to the complex field/disc interactions in the
pulsar-wind framework \cite{par17} was considered the most likely explanation
\cite{bog18,pap19,vel19}.  The reconnection of magnetic field lines threading the
disc, or of the donor star, could also explain the flares observed
at all wavelengths \cite{cam19}.

{\bf The variability -- } Simulations of the complex interaction between
the pulsar electromagnetic field and the disc plasma in a regime in
which they have comparable energy densities \cite{rom18,par17} helped  understand the
qualitative behaviour of tMSPs. However, they are  limited to
timescales of less than a second.  The long-term changes of state, the
{\it high}-{\it low} mode transitions and in general the flicker noise
variability lasting up to a few hours, were clearly out of reach.

\noindent The swings between accretion and rotation-powered activity over a few 
years are generally attributed to changes in the ram
pressure exerted by the matter captured by the gravitational field of the
NS. However, it remains unclear what ultimately causes the $\dot{M}$
to vary. A viscous disc instability is usually invoked to explain the
transient behaviour of dwarf novae and X-ray binaries
\cite{las01}. TMSPs challenged this interpretation since two very
different accretion disc states were realised, a bright and a {\it
  sub-luminous} one.  An advection dominated accretion flow may
characterise the fainter state, and a more typical geometrically
thin/optically thick disc the former. However, it is puzzling that
{\igr} (and possibly XMM J174457-2850.3) was able to show both states just
a few years apart.  Also, it is not yet determined whether a disc-like
flow manages to survive the pulsar wind pressure in the
rotation-powered state (see Sec.~\ref{sec:radiopulsarmodels}).
Variations in the mass-loss rate from the donor are unlikely driven by
changes of its radius since the timescales involved are much longer
even when high energy irradiation of the donor is considered. The magnetic activity of the secondary excited by its fast
orbital-locked rotation is instead an intriguing possibility to drive surges
of the mass transfer rate needed to squeeze the pulsar wind and start
the formation of an accretion disc. 

\noindent The driver of the recurring {\it high}-{\it low} mode transitions so
frequently observed in the {\it sub-luminous} state, despite a
chaotic and unstable disc/wind interaction, is also yet to determine.
A flicker noise power-law extending to very low frequencies
($10^{-3}-10^{-4}$~Hz) characterised the X-ray power density spectra
of both the bright outburst of {\igr} and the {\it sub-luminous}
states. Similar spectra have been sometimes observed from a few black
hole binaries in the soft state, while the noise power density of
AMXPs becomes flat below 0.1--1~Hz \cite{vst05}. The fluctuations of the
mass inflow rate in the outer disc. where the viscous time-scales are
long, are
usually invoked to explain such spectra \cite{lyu97}; these fluctuations must be able to propagate to the inner disc regions to produce the observed X-ray variability. In this
framework, the {\it high}-{\it low} mode transitions occurring in  tMSPs in the
{\it sub-luminous} state (but also the two intensity states observed from {\igr}
in outburst, \cite{fer14}) could reflect how the $\dot{M}$ variations
introduced in the outer disc eventually force the system into two well-defined luminosity states, related to the different regimes (accretion, propeller, radio pulsar) introduced
earlier.

\section{Conclusions}

A decade of observations of tMSPs demonstrated that variations in the
mass inflow rate lead to very different states in quick
succession. This unique property has allowed us to study  the complex interaction between the in-flowing plasma 
and the pulsar electromagnetic field in different regimes.

\noindent Multi-wavelength and high-temporal resolution simultaneous
observations of tMSPs have 
been  crucial to glimpse the physical
processes lying behind the  {\it sub-luminous} disc state. Much has to 
be done yet, especially to understand the nature of the modes and the
flares that characterise this state. For instance, establishing whether the X-ray and optical pulsations originate just outside the light cylinder, implying that the pulsar wind is terminated a few km away, would have important consequences to confirm the striped wind configuration and measure how it interacts with the surrounding matter.
On one hand, searches for more optical MSPs in either accretion or rotation-powered systems
will assess the nature of optical pulsations. On the other, MHD
simulations will help investigate the properties of the disc/wind intrabinary shock. In this regard, studies of
tMSPs could play an important role as a benchmark for the theories
that assume that a millisecond magnetar form after a double NS merger
and powers a short gamma-ray burst.

\noindent Searches for new systems associated with gamma-ray sources and displaying a peculiar X-ray and optical behaviour are intensively ongoing.  Together with the  
monitoring of redbacks and AMXPs, this  will likely increase the number of
confirmed tMSPs. This will be crucial to assess
whether all the MSPs in tight binaries are potentially transitional,
or other properties (e.g. magnetic activity of the secondary, magnetic inclination) are required to yield the state transitions.
Enlarging the sample has important consequences in the understanding of the population of short orbital period MSP binaries,  rather than restricting to the case by case, as done so far.

\section{Acknowledgements}
We acknowledge financial support  
from the Italian Space Agency (ASI) and National Institute for Astrophysics (INAF) under agreements ASI-INAF I/037/12/0 and ASI-INAF n.2017-14-H.0 and from INAF ”Main streams”, Presidential Decree 43/2018 and ”SKA/CTA projects”, Presidential Decree N. 70/2016. We acknowledge fruitful collaboration with F.~Ambrosino, E.~Bozzo, L.~Burderi, M.~Burgay, S.~Campana, F.~Coti Zelati, P.~D'Avanzo, T.~Di Salvo, C. Ferrigno, A.~Ghedina, A.~Miraval Zanon, F.~Meddi, E.~Poretti, A.~Possenti, N.~Rea, A.~Sanna, L.~Stella, D.~F.~Torres, and many more with whom we have faced the challenge of tMSPs in the last decade.

\bibliographystyle{spphys.bst}
\bibliography{tmspbiblio}
\end{document}